%% file: main.tex
\documentclass[preprint]{elsarticle}

\usepackage{amssymb}
\usepackage[a4paper,portrait,margin=1in]{geometry}
\usepackage{pdfpages}
\usepackage{dsfont}
\usepackage{complexity}
\usepackage{amsmath}
\usepackage{centernot}
\usepackage{lmodern}
\usepackage{amssymb}
\usepackage{float}
\usepackage{calc}	
\usepackage{tabstackengine}	
\usepackage{colortbl}
\usepackage{wrapfig}
\usepackage{mathrsfs}
\usepackage{multirow}

\usepackage{tikz}
\usetikzlibrary{decorations.markings, svg.path, fpu, shadows}
\makeatletter
\tikzset{use fpu reciprocal/.code={%
\def\pgfmathreciprocal@##1{%
    \begingroup
    \pgfkeys{/pgf/fpu=true,/pgf/fpu/output format=fixed}%
    \pgfmathparse{1/##1}%
    \pgfmath@smuggleone\pgfmathresult
    \endgroup
}}}%
\makeatother

\usepackage{xcolor}
\usepackage{url}
\usepackage{svg}
\usepackage{paralist, enumerate}
\usepackage{refcount}
\usepackage{subcaption}

\usepackage[noend]{algpseudocode}
\usepackage{algorithm}
\algrenewcommand\textproc{}

\usepackage{mathtools}
\usepackage{enumitem}
\usepackage{changepage}
\usepackage{comment}
\usepackage{color}	
\usepackage{soul}
\usepackage{pgfgantt}
\usepackage{verbatim}
\usepackage[UKenglish]{isodate}
\usepackage{tcolorbox}
\usepackage[binary-units]{siunitx}
\usepackage{afterpage}
\usepackage{xspace}
\usepackage{multirow,bigdelim}

\usepackage{makecell}

\newcommand\Tstrut{\rule{0pt}{2.6ex}}         % = `top' strut
\newcommand\Bstrut{\rule[-0.9ex]{0pt}{0pt}}   % = `bottom' strut

\definecolor{mylinkcolor}{rgb}{0,0,0.8}
\usepackage{hyperref} 
\makeatletter
\tikzset{
  reset label anchor/.code={%
    \let\tikz@auto@anchor=\pgfutil@empty
    \def\tikz@anchor{#1}
  },
  reset label anchor/.default=center,
  every label/.append style={reset label anchor}
}
\makeatother
    
\makeatletter	
\newcommand{\StatexIndent}[1][3]{%	
  \setlength\@tempdima{\algorithmicindent}%	
  \Statex\hskip\dimexpr#1\@tempdima\relax}	
\algdef{S}[WHILE]{WhileNoDo}[1]{\algorithmicwhile\ #1}%	
\makeatother	
	
\algnewcommand{\LeftComment}[1]{\Statex \(\triangleright\) #1}
\algnewcommand{\LineComment}[1]{\State \(\triangleright\) #1}

\usepackage{IEEEtrantools,stackengine}

\usepackage{amsthm}

\usepackage{thmtools}
\declaretheorem[name=Theorem]{thm}
\declaretheorem[name=Lemma]{lem}
\declaretheorem[name=Proposition]{prop}

\declaretheorem[name=Observation]{observation}
\declaretheorem[name=Corollary]{cor}

\newcommand{\algorithmfont}[1]{\textsf{#1}}

\def\mistfvariant/{MIS-3-TF}
\def\maxtwosatthreeshort/{M2S3}
\def\maxtwosatthree/{Max $2\text{SAT}_{\leq 3}$}
\def\porschenxsatvariant/{$\text{X3SAT}_{=3}^{+}$}

\newcommand{\vdkr}{\texorpdfstring{VDK\textsubscript{$r$}}{VDKr}\xspace}
\newcommand{\edkr}{\texorpdfstring{EDK\textsubscript{$r$}}{EDKr}\xspace}
\newcommand{\vdktwo}{\texorpdfstring{VDK\textsubscript{$2$}}{VDK2}\xspace}
\newcommand{\edktwo}{\texorpdfstring{EDK\textsubscript{$2$}}{EDK2}\xspace}
\newcommand{\vdkthree}{\texorpdfstring{VDK\textsubscript{$3$}}{VDK3}\xspace}
\newcommand{\edkthree}{\texorpdfstring{EDK\textsubscript{$3$}}{EDK3}\xspace}

\newcommand{\edkfour}{\texorpdfstring{EDK\textsubscript{$4$}}{EDK4}\xspace}

\newcommand{\edkfive}{\texorpdfstring{EDK\textsubscript{$5$}}{EDK5}\xspace}

\begin{document}
\begin{frontmatter}

\begin{abstract}
\input{abstract}
\end{abstract}


\title{Packing \texorpdfstring{$K_r$}{Kr}s in bounded degree graphs}%\tnoteref{t1}}
%\tnotetext[t1]{This work was supported by the Engineering and Physical Sciences Research Council (grant numbers EP/R513222/1 and EP/X013618/1).}

\author[gla]{Michael McKay\texorpdfstring{\corref{cor1}}{}}
\ead{mikemckay2203@gmail.com}
\author[gla]{David Manlove}
\ead{david.manlove@glasgow.ac.uk}

\cortext[cor1]{Corresponding author}
\address[gla]{School of Computing Science, University of Glasgow, UK}

\date{Received: date / Accepted: date}
\end{frontmatter}

\section{Introduction}
\label{sec:krpacking_intro}
\input{sections/intro}

\section{Linear-time solvability}
\label{sec:krpacking_lineartime}
\input{sections/linear_time_solvability}

\section{Polynomial-time solvability}
\label{sec:krpacking_ptime}
\input{sections/poly_time_solvability}

\section{\texorpdfstring{$\APX$}{APX}-hardness}
\label{sec:krpacking_apxhardness}
\input{sections/hardness}

\section{Conclusion and future work}
\label{sec:krpacking_conclusion}
\input{sections/conclusion}

\section{Acknowledgements}
\input{sections/funding}

\bibliographystyle{elsarticle-num} 
\bibliography{michael}

\end{document}

%% file: abstract.tex
We study the problem of finding a maximum-cardinality set of $r$-cliques in an undirected graph of fixed maximum degree $\Delta$, subject to the cliques in that set being either vertex disjoint or edge disjoint. It is known for $r=3$ that the vertex-disjoint (edge-disjoint) problem is solvable in linear time if $\Delta=3$ ($\Delta=4$) but $\APX$-hard if $\Delta \geq 4$ ($\Delta \geq 5$).

We generalise these results to an arbitrary but fixed $r \geq 3$, and provide a complete complexity classification for both the vertex- and edge-disjoint variants in graphs of maximum degree $\Delta$.

Specifically, we show that the vertex-disjoint problem is solvable in linear time if $\Delta < 3r/2 - 1$, solvable in polynomial time if $\Delta < 5r/3 - 1$, and $\APX$-hard if $\Delta \geq \lceil 5r/3 \rceil - 1$. We also show that if $r\geq 6$ then the above implications also hold for the edge-disjoint problem. If $r \leq 5$, then the edge-disjoint problem is solvable in linear time if $\Delta < 3r/2 - 1$, solvable in polynomial time if $\Delta \leq 2r - 2$, and $\APX$-hard if $\Delta > 2r - 2$.

%% file: sections/intro.tex
\subsection{Background}
\label{sec:krpacking_background}

In this paper we consider two problems related to clique packings in undirected graphs. Both problems involve finding a maximum-cardinality set of $r$-cliques, in a given undirected graph, where $r$ is a fixed constant. We call such a set a \emph{$K_r$-packing}. In the first problem, which we call the \emph{Vertex-Disjoint $K_r$-Packing Problem} (\vdkr), the cliques in the $K_r$-packing must be pairwise vertex disjoint. In the second problem, which we call the \emph{Edge-Disjoint $K_r$-Packing Problem} (\edkr), the cliques in the $K_r$-packing must be pairwise edge disjoint. Note that in both problems $r$ is a fixed constant and does not form part of the problem input. If $r$ is not fixed then both problems generalise the well-studied problem of finding a clique of a given size~\cite{GJ79}. Note also that if a vertex-disjoint $K_r$-packing has cardinality $|V|/r$ then we refer to it as a \emph{$K_r$-factor}~\cite{guruswami_k_2001}.

Most existing research concerning vertex- and edge-disjoint $K_r$-packings relates to either special or more general cases. For example, a special case of \vdkr is \vdktwo, also known as \emph{Maximum Cardinality Matching}. Maximum Cardinality Matching is is a central problem of graph theory and algorithmics~\cite{LP09}. A classical result of Edmonds~\cite{Edm65} is that a maximum-cardinality matching can be found in polynomial time. Conversely, \edktwo is trivial.

\vdkthree and \edkthree have been the subject of much research. In particular, \vdkthree is closely associated with the decision problem known as \emph{Partition Into Triangles} (PIT)~\cite{GJ79}, which asks whether a given undirected graph contains a $K_3$-factor. Karp~\cite{Karp75} noted in 1975 that PIT is $\NP$-complete. In 2002, Caprara and Rizzi~\cite{caprara_packing_2002} considered \vdkthree and \edkthree in graphs of a fixed maximum degree $\Delta$. They showed that \vdkthree is solvable in polynomial time if $\Delta \leq 3$ and $\APX$-hard even when $\Delta = 4$, and \edkthree is solvable in polynomial time if $\Delta \leq 4$ and $\APX$-hard even when $\Delta = 5$. They also showed that \vdkthree is $\NP$-hard for planar graphs even when $\Delta = 4$ and \edkthree is $\NP$-hard for planar graphs even when $\Delta = 5$. In their paper, Caprara and Rizzi~\cite{caprara_packing_2002} referenced a well-known approximation algorithm of Hurkens and Schrijver~\cite{hs89} for a more general type of packing problem. They noted that this algorithm leads to, for any fixed constant $\varepsilon > 0$, a $(3/2 + \varepsilon)$-approximation algorithm for \vdkthree and \edkthree. In 2013, van Rooij et al.~\cite{van_rooij_partition_2013} established an equivalence between \vdkthree when $\Delta = 4$ and \emph{Exact 3-Satisfiability} (X3SAT). They used this equivalence to devise an $O(1.02220^n)$-time algorithm for PIT when $\Delta = 4$. 

A well-studied generalisation of \vdkr involves finding in a given undirected graph $H$ a maximum-cardinality set of vertex-disjoint subgraphs where each subgraph is isomorphic to some fixed graph $G$. Such a set is known as a \emph{$G$-packing}. A $G$-packing is a \emph{$G$-factor} if the packing has cardinality $|V(H)|/|V(G)|$, where $V(H)$ is the set of vertices in $H$ and $V(G)$ is the set of vertices in $G$. In 1978, Kirkpatrick and Hell~\cite{KH78} showed that if $G$ contains a component with three or more vertices then it is $\NP$-complete to decide whether a given undirected graph contains a $G$-factor. In 1983, Kirkpatrick and Hell~\cite{KH83} surveyed previous work on $G$-packing, and also consider a further generalisation to so-called \emph{$\mathscr{G}$-packing}, where $\mathscr{G}$ is a fixed set of graphs and any subgraph in the $\mathscr{G}$-packing must be isomorphic to some element of $\mathscr{G}$. A survey of research involving $G$-packing can be found in a paper of Yuster~\cite{Yuster07} published in 2007.

Another interesting generalisation of \vdkr involves $\mathscr{G}$-packings such that $\mathscr{G}$ is a set of cliques. Of course, any $K_r$-packing problem can be seen as a $\mathscr{G}$-packing problem in which $\mathscr{G}=\{ K_r \}$. In 1984, Kirkpatrick and Hell~\cite{KH84Gpackings} presented polynomial-time algorithms for any case of the vertex-disjoint $\mathscr{G}$-packing problem in which $\mathscr{G} \subseteq \{ K_2, K_3, \dots \}$ and $\mathscr{G}$ contains $K_2$. In the same paper they also showed that the decision version of the $\mathscr{G}$-packing problem is $\NP$-complete for any set $\mathscr{G}$ where $\mathscr{G} \subseteq \{ K_3, K_4, \dots \}$. In 2008, Chataigner et al.~\cite{CHATAIGNER20091396} studied a related optimisation problem in which $\mathscr{G} = \{ K_2, K_3, \dots, K_r \}$ and the goal is to maximise the number of edges covered by such a $\mathscr{G}$-packing. They showed that if $r=3$ then this problem is $\APX$-complete, even when the input graph has fixed maximum degree $4$. They also presented new approximation algorithms, which in some cases improve on approximation ratios obtained via the previously-mentioned result of Hurkens and Schrijver~\cite{hs89}.

Compared to their various special and more general cases, \vdkr and \edkr as we have described them here appear to have received less attention in the literature. In 1998, Dahlhaus and Karpinski~\cite{DAHLHAUS199879} considered \vdktwo and \vdkr in chordal and strongly chordal graphs. They showed that a $K_r$-factor can be found in polynomial time in a given chordal or strongly chordal graph, if it exists. They remarked that if $r \geq 4$ then the decision version of \vdkr is $\NP$-complete for split graphs (a subset of chordal graphs), but left open the complexity for split graphs and chordal graphs when $r = 3$. Later, in 2001, Guruswami et al.~\cite{guruswami_k_2001} also considered \vdkr in relation to restricted classes of graphs, and resolved the open question of Dahlhaus and Karpinski. Guruswami et al.\ showed that if $r \geq 3$ then the decision version of \vdkr is $\NP$-complete for chordal graphs, planar graphs (only for $r = 3$ and $r = 4$), line graphs, and total graphs. They also described polynomial-time algorithms for \vdkthree in split graphs, the $K_r$-factor decision problem in split graphs, and \vdkr in cographs (also known as $P_4$-free graphs). They noted that this completely characterised the complexity of \vdkr for split graphs. The algorithm of Guruswami et al.\ for cographs was later extended by Pedrotti and de Mello~\cite{KrPackingP4sparse} for so-called $P_4$-sparse graphs.

The approximability of \vdkr and \edkr has also been studied. It is straightforward to apply the previously-discussed result of Hurkens and Schrijver~\cite{hs89} to show that there exists a polynomial-time $(r/2 + \varepsilon)$-approximation algorithm for \vdkr and \edkr, for any fixed constant $\varepsilon > 0$. In 2005, Mani\'c and Wakabayashi~\cite{Eurocomb05} described approximation algorithms that improve on this approximation ratio in the restricted cases of \vdkthree when $\Delta=4$, and \edkthree when $\Delta=5$. They also presented a linear-time algorithm for \vdkthree on indifference graphs.

% https://arxiv.org/pdf/1412.1774.pdf is useful here, Treglown

From the converse perspective of graphs with a fixed minimum degree, a classical result of Hajnal and Szemer\`edi~\cite{hajnal_szemeredi} is that a given undirected graph $G=(V, E)$ contains a $K_r$-factor if it has a minimum degree greater than or equal to $(1 - 1/r)|V|$. Kierstead and Kostochka~\cite{KIERSTEAD2008} later generalised this result to show that, in this case, such a packing can be constructed in polynomial time. Subsequent research has explored more general conditions for the existence of $K_r$-factors and $G$-factors~\cite{TreglownThesis,BKT13}.

\subsection{Our contribution}

Caprara and Rizzi~\cite{caprara_packing_2002} showed that \vdkthree is solvable in polynomial time if $\Delta=3$ and $\APX$-hard if $\Delta=4$; and \edkthree is solvable in polynomial time if $\Delta=4$ and $\APX$-hard if $\Delta=5$. In this paper we generalise their results and provide a full classification of the complexity of \vdkr and \edkr for any $\Delta \geq 1$ and any fixed $r \geq 3$. This classification in shown in Table~\ref{tab:krpacking_results}.

\begin{table}[ht]
\centering
\begin{tabular}{clp{0.0cm}lp{0.0cm}l}\noalign{\hrule}
\Tstrut\vspace*{0.2em}
& \multicolumn{3}{c}{is solvable in} & \multicolumn{2}{c}{\multirow{2}{*}{is $\APX$-hard if}} \\
& \multicolumn{1}{c}{linear time if} & \multicolumn{2}{c}{polynomial time if} &  \\
\hline\multirow{1}{*}{\vdkr}& $\Delta < 3r/2 - 1$ & & $\Delta < 5r/3 - 1$ & & $\Delta \geq \lceil 5r/3 \rceil - 1$\Tstrut\\[0.4em]
\multirow{2}{*}{\edkr}& \multirow{2}{*}{$\Delta < 3r/2 - 1$} & \ldelim\{{2}{*} & $\Delta \leq 2r - 2$ if $r \leq 5$ &  \ldelim\{{2}{*} &  $\Delta > 2r - 2$ if $r \leq 5$\\
& & & $\Delta < 5r/3 - 1$ otherwise & & $\Delta \geq \lceil 5r/3 \rceil - 1$ otherwise
\vspace*{0.2em}
\Bstrut\\\noalign{\hrule}
\end{tabular}
\caption{Our complexity results for \vdkr and \edkr where $r \geq 3$}
\label{tab:krpacking_results}
\end{table}

In the next section, Section~\ref{sec:krpacking_prelims}, we define some additional notation and make an observation on the coincidence of vertex- and edge-disjoint $K_r$-packings. In Section~\ref{sec:krpacking_lineartime} we consider the case when $\Delta < 3r/2 - 1$. We show that in this case any maximal vertex- or edge-disjoint $K_r$-packing is also maximum, and devise a linear-time algorithm for both \vdkr and \edkr in this setting. In Section~\ref{sec:krpacking_ptime} we present our algorithmic results, which show that \vdkr can be solved in polynomial time if $\Delta < 5r/3 - 1$; and \edkr can be solved in polynomial time if either $3 \leq r\leq 5$ and $\Delta \leq 2r - 2$, or $r\geq 6$ and $\Delta < 5r/3 - 1$. In Section~\ref{sec:krpacking_apxhardness} we show that our algorithmic results are in a sense best possible, unless $\P \neq \NP$. Specifically, we show that \vdkr is $\APX$-hard if $\Delta \geq \lceil 5r/3 \rceil - 1$; and \edkr is $\APX$-hard if either $3 \leq r \leq 5$ and $\Delta > 2r - 2$, or $r \geq 6$ and $\Delta \geq \lceil 5r/3 \rceil - 1$. In other words, we prove that there exist fixed constants $\varepsilon > 1$ and $\varepsilon' > 1$ such that no polynomial-time $\varepsilon$-approximation algorithm exists for \vdkr if $\Delta \geq \lceil 5r/3 \rceil - 1$; and no polynomial-time $\varepsilon'$-approximation algorithm exists for \edkr if either $3 \leq r \leq 5$ and $\Delta > 2r - 2$, or $r \geq 6$ and $\Delta \geq \lceil 5r/3 \rceil - 1$. In Section~\ref{sec:krpacking_conclusion} we recap our results and consider directions for future work.

\subsection{Preliminaries}
\label{sec:krpacking_prelims}
\input{sections/prelims}

%% file: sections/prelims.tex
In this section we clarify our notation and terminology and make a preliminary observation.

Let $G = (V, E)$ be a simple undirected graph. For any vertex $v$ in $V$ let the \emph{open neighbourhood} of $v$, denoted $N_{G}(v)$, be the set of vertices adjacent to $v$ in $G$ and let the \emph{closed neighbourhood} of $v$, denoted $N_{G}[v]$, be $N_{G}(v) \cup \{ v \}$. 
We denote by $\deg_{G}(v) = |N_{G}(v)|$ the \emph{degree} of $v$ in $G$ and by $\Delta(G) = \max_{v\in V} \deg_{G}(v)$ the \emph{maximum degree} of $G$. If the graph in question is clear from context then we just write $\Delta$. For any subset of vertices $U \subseteq V$, we denote by $G[U]$ the subgraph of $G$ induced by $U$. For any vertex $v_1 \in V$, if $N_{G}(v_1)$ contains three vertices $v_2, v_3, v_4$ that are an independent set then we say that the subgraph induced by $\{ v_1, v_2, v_3, v_4 \}$ in $G$ is a \emph{claw}. If no induced subgraph of $G$ is a claw then we say that $G$ is \emph{claw-free}~\cite{MINTY1980284}.

We write \emph{$K_r$} to mean a clique of size $r$, for some integer $r \geq 1$. Let $K_r^G$ be the set of $K_r$s in $G$. We say that a set $T$ is a \emph{$K_r$-packing} in $G$ if $T\subseteq K_r^G$. 
% The \emph{cardinality} of a $K_r$-packing is the number of $K_r$s that it contains. 
We say that a $K_r$-packing $T$ is \emph{vertex disjoint} if any two $K_r$s in $T$ have no vertex in common, and is \emph{edge disjoint} if any two $K_r$s in $T$ intersect by at most one vertex. The \emph{Vertex-Disjoint $K_r$-Packing Problem} (\vdkr) is the following optimisation problem: given a simple undirected graph $G$, find a vertex-disjoint $K_r$-packing of maximum cardinality. The \emph{Edge-Disjoint $K_r$-Packing Problem} (\edkr) is defined analogously.

For any maximisation problem $P$, instance $I$ of $P$, and feasible solution $S$ of $I$, let $\textrm{m}_{P}(I, S)$ denote the measure of $S$. Let $\textrm{opt}_{P}(I) = \max_{S \in \mathcal{F}(I)} \textrm{m}_{P}(I, S)$, where $\mathcal{F}(I)$ is the set of feasible solutions of $I$.

For technical purposes we define the \emph{$K_r$-vertex intersection graph} $\mathcal{K}_r^G = (K_r^G, E_{\mathcal{K}_r^G})$ of $G$, where $\{ U, W \} \in E_{\mathcal{K}_r^G}$ if and only if $|U \cap W| \geq 1$ for any $U, W \in K_r^G$. Similarly, we define the \emph{$K_r$-edge intersection graph} ${\mathcal{K}'}_r^G = (K_r^G, E_{{\mathcal{K}'}_r^G})$ of $G$ in which $\{ U, W \} \in E_{{\mathcal{K}'}_r^G}$ if $|U \cap W| \geq 2$ for any $U, W \in K_r^G$. We now make a preliminary observation.

\begin{observation}
\label{obs:krpacking_edkr_is_also_vdkr}
If $\Delta < 2r - 2$ then any edge-disjoint $K_r$-packing is also vertex disjoint.
\end{observation}
\begin{proof}
Any two $K_r$s that intersect by at least one vertex must in fact intersect by at least two vertices, since otherwise that vertex has degree at least $2r - 2$.
\end{proof}

%% file: sections/linear_time_solvability.tex
In this section we present an algorithm that can solve \vdkr and \edkr in linear time if $\Delta < 3r/2 - 1$. This algorithm generalises an algorithm of van Rooij et al.~\cite{van_rooij_partition_2013} that can solve \vdkthree in linear time if $\Delta \leq 3$.

The key insight behind this algorithm is that if $\Delta < 3r/2 - 1$ then any maximal vertex-disjoint $K_r$-packing is also a maximum vertex-disjoint $K_r$-packing. The proof of this is stated below in Theorem~\ref{thm:krpacking_r_maximal_is_maximum}, which we prove using a sequence of lemmas. In what follows, suppose $G=(V,E)$ is a simple undirected graph where $\Delta(G) < 3r/2 - 1$.

\begin{lem}
\label{lem:krpacking_r_lem1}
For any $U_i, U_j \in K_r^G$, if $\{ U_i, U_j \} \in E_{\mathcal{K}_r^G}$ then $|U_i \cap U_j| > r/2$.
\end{lem}
\begin{proof}
Consider some $U_i, U_j \in K_r^G$ where $\{ U_1, U_2 \} \in E_{\mathcal{K}_r^G}$. By the definition of the $K_r$-vertex intersection graph $\mathcal{K}_r^G$, there exists some vertex $u \in V$ where $u \in U_i \cap U_j$. Since $3r/2 - 1 > \Delta(G) \geq \deg_{G}(u) \geq |U_i \cup U_j| - 1 = |U_i| + |U_j| - |U_i \cap U_j| - 1 = 2r - |U_i \cap U_j| - 1$ it follows that $|U_i \cap U_j| > 2r - 3r/2 = r/2$.
\end{proof}

\begin{lem}
\label{lem:krpacking_r_lem3}
$\mathcal{K}_r^G$ is a disjoint union of cliques (i.e.\ a cluster graph~\cite{clustergraphcitation}).
\end{lem}
\begin{proof}
It suffices to show that if there exists three sets $U_i, U_j, U_k$ in $K_r^G$ where $\{ U_i, U_j \} \in E_{\mathcal{K}_r^G}$ and $\{ U_j, U_k \} \in E_{\mathcal{K}_r^G}$, then $\{ U_i, U_k \} \in E_{\mathcal{K}_r^G}$. Consider some such $U_i, U_j, U_k \in K_r^G$. Since $\{ U_i, U_j \} \in E_{\mathcal{K}_r^G}$ and $\{ U_j, U_k \} \in E_{\mathcal{K}_r^G}$,  by Lemma~\ref{lem:krpacking_r_lem1} it must be that $|U_i \cap U_j| > r/2$ and $|U_j \cap U_k|>r/2$. Since $|U_j|=r$ it follows that $|U_i \cap U_k| > 0$ and thus that $\{ U_i, U_k \} \in E_{\mathcal{K}_r^G}$.
\end{proof}

\begin{thm}
\label{thm:krpacking_r_maximal_is_maximum}
If $T$ is a maximal vertex-disjoint $K_r$-packing then $T$ is a maximum vertex-disjoint $K_r$-packing.
\end{thm}
\begin{proof}
Suppose $T$ is a maximal vertex-disjoint $K_r$-packing in $G$, which by definition corresponds to a maximal independent set in $\mathcal{K}_r^G$. Since $\mathcal{K}_r^G$ is the disjoint union of cliques (by Lemma~\ref{lem:krpacking_r_lem3}), any two maximal independent sets in $\mathcal{K}_r^G$ have the same cardinality, so $T$ is also maximum.
\end{proof}

We have shown in Theorem~\ref{thm:krpacking_r_maximal_is_maximum} that any maximal vertex-disjoint $K_r$-packing is also a maximum vertex-disjoint $K_r$-packing. It follows immediately that \vdkr can be solved in $O(|V|^r)$ time by constructing the $K_r$-vertex intersection graph $\mathcal{K}_r^G$ and greedily selecting an independent set. In fact, the explicit construction of $\mathcal{K}_r^G$ can be avoided by exploring $G$ and greedily selecting $K_r$s. We present Algorithm~\algorithmfont{greedyCliques}, shown in Algorithm \ref{alg:krpacking_r}, and show that it requires $O(|V|)$ time.

\input{algorithms/greedycliques}

\begin{lem}
\label{lem:krpacking_lineartimealgorunningtime}
Algorithm~\algorithmfont{greedyCliques} requires $O(|V|)$ time.
\end{lem}
\begin{proof}
In any iteration of the outermost while loop, either a single vertex $v$ or a non-empty set of vertices $K$ is removed from $G$. It follows that the algorithm terminates after at most $|V|$ iterations of this loop. It remains to show that one iteration of this loop can be performed in constant time.

In each iteration, either $\deg_{G}(v) \geq r-1$ or $\deg_{G}(v) < r - 1$. Computing $\deg_{G}(v)$ requires $O(r)$ time, since $\Delta < 3r/2 - 1$.  
Consider the first branch of the outermost if statement. There are $\binom{|N_G(v)|}{r - 1} \leq \binom{\Delta}{r - 1} < \binom{3r/2-1}{r-1}=O(2^r)$ iterations of the for loop. In each iteration, the algorithm tests if $G[W]$ contains $\binom{r-1}{2}$ edges. This can be performed in $O(r^2)$ time. Removing $K$ from $G$ and adding $K$ to $T$, if $K\neq \varnothing$, can be done in $O(r^2)$ time. In both the else branch in which $K=\varnothing$ and the second branch of the outermost if statement, $v$ can be removed from $G$ in $O(r)$ time.
\end{proof}

\begin{thm}
\label{thm:krpacking_vdkr_3r2minus1}
If $\Delta(G) < 3r/2 - 1$ then \vdkr can be solved in linear time.
\end{thm}
\begin{proof}
By Lemma~\ref{lem:krpacking_lineartimealgorunningtime}, Algorithm~\algorithmfont{greedyCliques} terminates in $O(2^r |V|)$ time. By Theorem~\ref{thm:krpacking_r_maximal_is_maximum}, it suffices to show that this algorithm returns a set $T$ that is a maximal vertex-disjoint $K_r$-packing in $G$. Suppose $K'$ is an arbitrary $K_r$ in $G$. We show that either $K'$ is added to $T$ or at least one vertex in $K'$ belongs to some other $K_r$ in $T$. By the pseudocode, the algorithm removes at least one vertex in each iteration of the while loop, which ends once there are no remaining vertices. Consider the first iteration of the while loop in which any vertex $v$ in $K'$ is identified and removed. Let $G'$ be the value of $G$ at the beginning of this iteration. By definition, at this point every vertex in $K'$ is present in $G'$, including $v$. Since $\deg_{G'}(v) \geq r - 1$, it must be that $v$ was not deleted from $G'$ by the second branch of the outermost if statement. Similarly, $v$ cannot have been deleted from $G'$ by the second branch of the innermost if statement, since $v$ belongs to $K'$, which is a clique of size $r$ in $G'$. The only possibility is that $v$ was deleted from $G'$ as a result of $v$ being part of some $K_r$ in $G'$, which was at some point added to $T$.
\end{proof}

\begin{cor}
\label{cor:krpacking_edkr_3r2minus2}
If $\Delta(G) < 3r/2 - 1$ then \edkr can be solved in linear time.
\end{cor}
\begin{proof}
If $r \leq 2$ then \edkr is trivial. If $r \geq 3$ then it must be that $\Delta < 3r/2 - 1 < 2r - 2$, so by Observation~\ref{obs:krpacking_edkr_is_also_vdkr} any edge-disjoint $K_r$-packing is also vertex disjoint. It follows that any maximum vertex-disjoint $K_r$-packing returned by Algorithm~\algorithmfont{greedyCliques} is also a maximum edge-disjoint $K_r$-packing. 
\end{proof}

%% file: algorithms/greedycliques.tex
\begin{algorithm}
\textbf{Input:} a fixed integer $r\geq 1$ and a simple undirected graph $G=(V, E)$ where $\Delta(G) < 3r/2 - 1$\\
\textbf{Output:} a maximum $K_r$-packing $T$
\smallskip
\begin{algorithmic}
\caption{Algorithm~\algorithmfont{greedyCliques}\label{alg:krpacking_r}} 
\State $T\gets\varnothing$
\While{$|V| > 0$}
    \State $v\gets \text{any vertex in }V$
    \If{$\deg_{G}(v) \geq r - 1$}
        \State $K\gets \varnothing$
        \For{each subset $W$ of size $r-1$ of $N_{G}(v)$}
            \If{$G[W]$ has $\binom{r-1}{2}$ edges}
                \LineComment{$W$ must be a clique of size $r-1$ in $G$}
                \State $K \gets W \cup \{ v \}$
            \EndIf
            \State \textbf{end if}
        \EndFor
        \State \textbf{end for}
        \smallskip
        
        \If{$K\neq\varnothing$}
            \State $G \gets G[V \setminus K]$
            \State $T \gets T \cup \{ \{ K \} \}$
        \Else
            \State $G \gets G[V \setminus \{ v \}]$
        \EndIf
        \State \textbf{end if}
    \Else
        \State $G \gets G[V \setminus \{ v \}]$
    \EndIf
    \State \textbf{end if}
\EndWhile
\State \textbf{end while}
\smallskip

\State \Return $T$
\end{algorithmic}
\end{algorithm}

%% file: sections/poly_time_solvability.tex
\subsection{Vertex-disjoint \texorpdfstring{$K_r$}{Kr}-packing} 
\label{sec:krpacking_vdkr_polytimesolvability}

In this section we consider \vdkr. We show that \vdkr is solvable in polynomial time if $\Delta < 5r/3 - 1$. The proof involves finding an independent set in the $K_r$-vertex intersection graph $\mathcal{K}_r^G$. We build on the technique of Caprara and Rizzi~\cite{caprara_packing_2002} and first show that if $\Delta(G) < 5r/3 - 1$ then $\mathcal{K}_r^G$ is claw-free. It follows that a maximum independent set in $\mathcal{K}_r^G$ can be found in polynomial time~\cite{MINTY1980284,SBIHI198053}, which corresponds directly to a maximum vertex-disjoint $K_r$-packing.

There is an evident relationship between packing problems and independent sets in intersection graphs~\cite{Eurocomb05,Hal95,kann91}. In his paper on claw-free graphs, Minty~\cite{MINTY1980284} remarked that an algorithm to find a maximum cardinality matching (i.e.\ solve \vdktwo) can be used to find a maximum independent set in a line graph (i.e\ a $K_2$-vertex intersection graph). Here, like Caprara and Rizzi~\cite{caprara_packing_2002}, we make use of the converse relationship and show that if the corresponding intersection graph is claw-free then \vdkr and \edkr can be solved in polynomial time.

In what follows, suppose $G=(V,E)$ is an undirected graph where $\Delta(G) < {5r}/{3}-1$. In Lemma~\ref{lem:krpacking_polytime_overlap} we place a lower bound on the size of the intersection of any two $K_r$s in $G$ that intersect by at least one vertex.

\begin{lem}
\label{lem:krpacking_polytime_overlap}
$|U_i \cap U_j| > r/3$ for any $\{ U_i, U_j \} \in E_{\mathcal{K}_r^G}$.
\end{lem}
\begin{proof}
Consider some $\{ U_i, U_j \} \in E_{\mathcal{K}_r^G}$ and an arbitrary vertex $u_l \in |U_i \cap U_j|$. Now ${5r}/{3} - 1 > \Delta(G) \geq \deg_{G}(u_l) \geq |U_i \cup U_j| - 1 = 2r - |U_i \cap U_j| - 1$. Rearranging gives $|U_i \cap U_j| > r/3$.
\end{proof}

\begin{lem}
\label{lem:krpacking_vdkr_clawfree}
$\mathcal{K}_r^G$ is claw-free.
\end{lem}
\begin{proof}
Consider some $U_i, U_{j_1}, U_{j_2}, U_{j_3} \in K_r^G$ where $\{ U_i, U_{j_a} \} \in E_{\mathcal{K}_r^G}$ for each $a \in \{1, 2, 3\}$. By Lemma~\ref{lem:krpacking_polytime_overlap}, it must be that $|U_i \cap U_{j_1}| > r/3$, $|U_i \cap U_{j_2}| > r/3$, and $|U_i \cap U_{j_3}| > r/3$. Since $|U_i|=r$ it follows by the pigeonhole principle that either $U_{j_1}$ intersects $U_{j_2}$, $U_{j_2}$ intersects $U_{j_3}$, or $U_{j_1}$ intersects $U_{j_3}$, so the subgraph induced by $\{ U_i, U_{j_1}, U_{j_2}, U_{j_3} \}$ is not a claw in $\mathcal{K}_r^G$.
\end{proof}

\begin{thm}
\label{thm:krpacking_vdkr_polytime_5r3minus1}
If $\Delta(G)<5r/3-1$ then \vdkr can be solved in polynomial time.
\end{thm}
\begin{proof}
First, construct the $K_r$-vertex intersection graph $\mathcal{K}_r^G = (K_r^G, E_{\mathcal{K}_r^G})$. The set $K_r^G$ can be constructed in $O(\binom{|V|}{r})=O(|V|^r)$ by considering every possible set of $r$ vertices in $V$. The set $E_{\mathcal{K}_r^G}$ can then be constructed in $O(|V|^{2r})$ time. Next, find a maximum independent set in $\mathcal{K}_r^G$, which can be done in polynomial time since $\mathcal{K}_r^G$ is claw-free (by Lemma~\ref{lem:krpacking_vdkr_clawfree})~\cite{SBIHI198053,MINTY1980284}.
\end{proof}

Given a claw-free graph in which each vertex has a real weight, there exists polynomial-time algorithms that can find an independent set of maximum total weight~\cite{Nakamura01,FOS11}. We remark that the the result shown in Theorem~\ref{thm:krpacking_vdkr_polytime_5r3minus1} can be generalised to a version of \vdkr in which vertices or edges have weights, and the goal is to find a $K_r$-packing of maximum total weight. 

\subsection{Edge-disjoint \texorpdfstring{$K_r$}{Kr}-packing} 

In this section we consider \edkr. Using Theorem~\ref{thm:krpacking_vdkr_polytime_5r3minus1} and Observation~\ref{obs:krpacking_edkr_is_also_vdkr}, it is straightforward to show that if $\Delta < {5r}/{3} - 1$ then \edkr can be solved in polynomial time. We state this result as Theorem~\ref{thm:krpacking_edkrgeq6_polytime}. In what follows, suppose $G$ is an undirected graph.

\begin{thm}
\label{thm:krpacking_edkrgeq6_polytime}
If $\Delta(G) < 5r/3 - 1$ then \edkr can be solved in polynomial time.
\end{thm}
\begin{proof}
If $\Delta(G) < 5r/3 - 1$ then we can find a maximum vertex-disjoint $K_r$-packing in polynomial time by Theorem~\ref{thm:krpacking_vdkr_polytime_5r3minus1}. Such a packing is also a maximum edge-disjoint $K_r$-packing, by Observation~\ref{obs:krpacking_edkr_is_also_vdkr}.
\end{proof}

We now show that this upper bound on $\Delta(G)$ can be increased if $r \in \{ 4, 5 \}$. The key insight in this case is that if $r \in \{ 4, 5 \}$ and $\Delta \leq 2r - 2$ then the $K_r$-edge intersection graph ${\mathcal{K}'}_r^G$ is claw-free. This is the same technique used by Caprara and Rizzi~\cite{caprara_packing_2002} to show that \edkthree is solvable in polynomial time when $\Delta(G) \leq 4$.

\begin{lem}
\label{lem:krpacking_edgedisjoint_r45}
If $r\in \{ 4, 5 \}$ and $\Delta(G) \leq 2r - 2$ then the $K_r$-edge intersection graph ${\mathcal{K}'}_r^G$ is claw-free.
\end{lem}
\begin{proof}
Consider some $U_i, U_{j_1}, U_{j_2}, U_{j_3} \in K_r^G$ where $\{ U_i, U_{j_a} \} \in E_{{\mathcal{K}'}_r^G}$ for each $a \in \{1, 2, 3\}$. Suppose for a contradiction that the subgraph induced by $\{ U_i, U_{j_1}, U_{j_2}, U_{j_3} \}$ in ${\mathcal{K}'}_r^G$ is a claw. By the definition of $E_{{\mathcal{K}'}_r^G}$, it must be that $|U_{j_a} \cap U_{j_b}| < 2$ for any $a$ and $b$ where $a \neq b$ and $a, b \in \{ 1, 2, 3 \}$. It must also be that $|U_i \cap U_{j_a}| \geq 2$ for each $a \in \{ 1, 2, 3 \}$. Since $|U_i| = r \leq 5$, assume without loss of generality that $|U_i \cap U_{j_1} \cap U_{j_2}| \geq 1$. Furthermore, it must be that $|U_i \cap U_{j_1} \cap U_{j_2}| = 1$, otherwise $|U_{j_1} \cap U_{j_2}| > 1$ which is a contradiction.  Let $v_r$ be the single vertex in $U_i \cap U_{j_1} \cap U_{j_2}$. Since $U_{j_1}$ and $U_{j_2}$ are $K_r$s in $G$ and $\deg_{G}(v_r) \leq 2r-2$ it must be that $N_{G}[v_r]=U_{j_1} \cup U_{j_2}$. Since $U_i$ is also a $K_r$ it follows that $U_i \subset (U_{j_1} \cup U_{j_2})$. 

Now consider $U_i \cap U_{j_1}$ and $U_i \cap U_{j_2}$. If $|U_i \cap U_{j_1}| + |U_i \cap U_{j_2}| \geq r+2$ then since $|U_i|=r$ it follows that $|U_{j_1} \cap U_{j_2}| \geq 2$ which is a contradiction. It follows that $|U_i \cap U_{j_1}| + |U_i \cap U_{j_2}| \leq r + 1$ and either $|U_i \cap U_{j_1}| \leq (r+1)/2$ or $|U_i \cap U_{j_2}| \leq (r+1)/2$. Assume without loss of generality that $|U_i \cap U_{j_1}| \leq (r+1)/2$. 

Now consider $U_{j_3}$. Since the subgraph induced by $\{ U_i, U_{j_1}, U_{j_2}, U_{j_3} \}$ in ${\mathcal{K}'}_r^G$ is a claw, it must be that $|U_{j_3} \cap U_{i}| \geq 2$, $|U_{j_3} \cap U_{j_1}| \leq 1$, and $|U_{j_3} \cap U_{j_2}| \leq 1$. Since $U_i \subset (U_{j_1} \cup U_{j_2})$, the only possibility is that $|U_{j_3} \cap U_i| = 2$, $|U_{j_3} \cap U_i \cap U_{j_2}| = 1$ and $|U_{j_3} \cap U_i \cap U_{j_1}| = 1$. 

Let $v_s$ be the single vertex in $U_{i} \cap U_{j_1} \cap U_{j_3}$. Since $U_i$, $U_{j_1}$, and $U_{j_3}$ are $K_r$s in $G$ it follows that $v_s$ is adjacent to every other vertex in $U_{j_3} \cup U_{i} \cup U_{j_1}$ so
\begin{align*}
    \deg_{G}(v_s) &\geq |U_i \cup U_{j_1} \cup U_{j_3}| - 1\\
    &= 3r - |U_i \cap U_{j_1}| - |U_{i} \cap U_{j_3}| - |U_{j_1} \cap U_{j_3}| + |U_i \cap U_{j_1} \cap U_{j_3}| - 1\enspace.
\end{align*}
Recall that since the subgraph induced by $\{ U_i, U_{j_1}, U_{j_2}, U_{j_3} \}$ in ${\mathcal{K}'}_r^G$ is a claw, $|U_{j_1} \cap U_{j_3}| \leq 1$. We deduced earlier that $|U_i \cap U_{j_1}| \leq (r+1)/2$, $|U_i \cap U_{j_3}| = 2$, and $|U_{j_3} \cap U_i \cap U_{j_1}| = 1$. Since $r \geq 4$ it follows that $\deg_{G}(v_s) \geq (5r-7)/2 > 2r - 2$, which is a contradiction.
\end{proof}

\begin{thm}
\label{thm:krpacking_edkr345_polytime}
If $r \leq 5$ and $\Delta(G) \leq 2r - 2$ then \edkr can be solved in polynomial time.
\end{thm}
\begin{proof}

If $r\leq 2$ then \edkr is trivial. Caprara and Rizzi \cite{caprara_packing_2002} prove the case when $r=3$ and $\Delta(G) \leq 4$. If $r \in \{ 4, 5 \}$ and $\Delta(G) \leq 2r - 2$ then by Lemma~\ref{lem:krpacking_edgedisjoint_r45}, the $K_r$-edge intersection graph ${\mathcal{K}'}_r^G$ is claw-free. It follows that a maximum edge-disjoint $K_r$-packing can be found in polynomial time by constructing ${\mathcal{K}'}_r^G$, in $O(|V|^{2r})$ time, and finding in it a maximum independent set, which can also be accomplished in polynomial time~\cite{SBIHI198053,MINTY1980284}.
\end{proof}

As we remarked for \vdkr in Section~\ref{sec:krpacking_vdkr_polytimesolvability},  it seems that the results shown in Theorems~\ref{thm:krpacking_edkrgeq6_polytime} and~\ref{thm:krpacking_edkr345_polytime} can be generalised to a version of \edkr in which vertices or edges have weights, and the goal is to find a $K_r$-packing of maximum total weight.

%% file: sections/hardness.tex
\subsection{Vertex-disjoint \texorpdfstring{$K_r$-packing}{Kr-packing}} 

We now show that if $r \geq 3$ then \vdkr is $\APX$-hard even when $\Delta \geq \lceil 5r/3 \rceil - 1$. In other words, for any $r \geq 3$ there exists some fixed constant $\varepsilon > 0$ such that no polynomial-time $\varepsilon$-approximation algorithm exists for \vdkr even when $\Delta \geq \lceil 5r/3 \rceil - 1$, unless $\P=\NP$.

We reduce from the problem of finding a Maximum Independent Set (MIS) in a graph $G=(V,E)$ that has maximum degree $3$ and is triangle-free. Berman and Karpinski~\cite{BK99} show that this optimisation problem, which we refer to as \mistfvariant/, is $\APX$-hard, notably providing an explicit lower bound on the approximation ratio (specifically, they showed that it is $\NP$-hard to approximate \mistfvariant/ within $140/139-\varepsilon$, for any fixed $\varepsilon > 0$).

The reduction from \mistfvariant/ is as follows. Our goal is to construct a new graph $G'=(V', E')$ where each $K_r$ in $G'$ corresponds to exactly one vertex in $V$ and each vertex in $V$ corresponds to exactly one $K_r$ in $G'$. For any two adjacent vertices in $G$, the intersection of the two corresponding $K_r$s in $G'$ will contain exactly $\lfloor r/3 \rfloor$ vertices. 

To do this, first construct a set of $|V|$ disjoint $K_r$s in $G'$, labelled $\mathcal{U} = \{ U_1, U_2, \dots, U_{|V|} \}$ where $U_i = \{ u_i^1, u_i^2, \dots, u_i^r \}$ for any $i$ where $1\leq i \leq |V|$. Next, consider each edge $\{ v_i, v_j \} \in E$. let $U'_i = \{ u_i^{a_1}, u_i^{a_2}, \dots, u_i^{a_{\lfloor r/3 \rfloor}} \}$ be any set of $\lfloor r/3 \rfloor$ vertices in $U_i$ with degree $r-1$ and $U'_j = \{ u_j^{b_1}, u_j^{b_2}, \dots, u_j^{b_{\lfloor r/3 \rfloor}} \}$ be any set of $\lfloor r/3 \rfloor$ vertices in $U_j$ with degree $r-1$. For each $q$ from $1$ to $\lfloor r/3 \rfloor$ inclusive, identify $u_i^{a_q}$ and $u_j^{b_q}$ to create a single vertex labelled $u_{ij}^{a_q}$. Label $U'_j = U'_i$ as $W_{ij}$. % Note that since $\Delta(G) = 3$, for any edge $\{ v_i, v_j \} \in E$ considered there must exist some set of $\lfloor r/3 \rfloor$ vertices in $U_i$ with degree $r-1$ and some set of $\lfloor r/3 \rfloor$ vertices in $U_j$ with degree $r-1$.

Finally, for each vertex $v_i \in V$ let $X_i$ be the set of (at least $r \bmod 3$) vertices in $U_i$ with degree $r-1$. Note that any vertex in $G'$ either belongs to some set $W_{ij}$ where $\{ v_i, v_j \} \in E$ or some set $X_i$ where $v_i \in V$.

We first show that the set of $K_r$s in $G'$ is $\mathcal{U}$.

\begin{lem}
\label{lem:krpacking_lissetofkrs}
$\mathcal{U} = K_r^{G'}$.
\end{lem}
\begin{proof}
% By construction, before any vertex identifications are made it must be that any $K_r$ in $G$ belongs to $\mathcal{U}$. Suppose for a contradiction that at some point in the reduction $u_i^{a_q}$ and $u_j^{b_q}$ are identified, after which there exists 
By definition, $\mathcal{U} \subseteq K_r^{G'}$ so it remains to show that each $K_r$ in $G'$ belongs to $\mathcal{U}$. Suppose $K$ is an arbitrary $K_r$ in $G'$. By definition, any vertex in any set $X_i$ has degree $r-1$ in $G'$ and thus belongs to exactly one $K_r$ in $G'$, namely $U_i$, which belongs to $K_r^{G'}$. It follows that each vertex in $K$ belongs to some set $W_{ij}$ where $1\leq i, j \leq |V|$. Since $|W_{ij}| = \lfloor r/3 \rfloor$ it must be that either there exist three sets $W_{i_1 j_1}, W_{i_2 j_2}, W_{i_3 j_3}$ where $1\leq i_1, i_2, \dots, j_3 \leq |V|$ and $K = W_{i_1 j_1} \cup W_{i_2 j_2} \cup W_{i_3 j_3}$, or there exist four or more sets $W_{i_1 j_1}, W_{i_2 j_2}, W_{i_3 j_3}, W_{i_4, j_4}, \dots$ where $1\leq i_1, i_2, \dots \leq |V|$ and $K$ contains at least one vertex in each set. In the latter case, we may assume without loss of generality that $\{ i_1, j_1 \} \cap \{ i_2, j_2 \} = \varnothing$. By the construction of $G'$ it follows that no edge exists between any vertex in $W_{i_1 j_1}$ and any vertex in $W_{i_2 j_2}$, which contradicts the supposition that $K$ is a $K_r$ in $G'$. It remains that there exist three sets $W_{i_1 j_1}, W_{i_2 j_2}, W_{i_3 j_3}$ where $K = W_{i_1 j_1} \cup W_{i_2 j_2} \cup W_{i_3 j_3}$ and $1\leq i_1, i_2, \dots, j_3 \leq |V|$.

By construction, the closed neighbourhood of any vertex in $W_{i_1 j_1}$ is $U_{i_1} \cup U_{j_1}$ so since $K$ is a $K_r$, without loss of generality assume that $i_1 \in \{ i_2, j_2 \}$ and $j_1 \in \{ i_3, j_3 \}$. A symmetric argument shows that $i_2 \in \{ i_1, j_1 \}$ and $j_2 \in \{ i_3, j_3 \}$, and $i_3 \in \{ i_1, j_1 \}$ and $j_3 \in \{ i_2, j_2 \}$. By symmetry, we need only consider the two cases, in which $i_1 = i_2 = i_3$ and in which $K = W_{i_1, i_2} \cup W_{i_2, i_3} \cup W_{i_3, i_1}$. In the former case, $K$ must be labelled $U_{i_1}$ and thus belongs to $K_r^G$. In the latter case, by the construction of $G'$ the three vertices $\{ v_{i_1}, v_{i_2}, v_{i_3} \}$ in $G$ form a triangle, which is a contradiction.
\end{proof}

\begin{lem}
\label{lem:krpacking_vdkr_reduction_degree}
$\Delta(G') = \lceil 5r/3 \rceil - 1$.
\end{lem}
\begin{proof}
By definition, any vertex in any set $X_i$ has degree $r-1$. Any vertex in any set $W_{ij}$ has degree $|U_i| + |U_j| - |W_{ij}| - 1 = 2r - \lfloor r/3 \rfloor - 1 = \lceil 5r/3 \rceil - 1$, since $r$ is an integer.
\end{proof}

\begin{thm}
\label{thm:krpacking_vdkr_apxhard}
If $r \geq 3$ and $\Delta \geq \lceil 5r/3 \rceil - 1$ then \vdkr is $\APX$-hard.
\end{thm}
\begin{proof}
We first show that a vertex-disjoint $K_r$-packing of size $B$ exists in $G'$ if and only if an independent set of size $B$ exists in $G$. By the design of the reduction, and Lemma~\ref{lem:krpacking_lissetofkrs}, any two $K_r$s in $K_r^{G'}$ that are not vertex disjoint in $G'$ correspond to two vertices in $G$ that are adjacent. Conversely, for any two vertices in $G$ that are adjacent, by the design of the reduction it must be that the two corresponding $K_r$s in $K_r^{G'}$ are not vertex disjoint in $G'$. 
It follows that, for any graph $G$, $\textrm{opt}_{\textrm{\mistfvariant/}}(G) = \textrm{opt}_{\textrm{\vdkr}}(G')$. Moreover, for any vertex-disjoint $K_r$-packing $T$ in $G'$, there exists a corresponding independent set $S$ in $G$ where $|S|=|T|$, and thus that $\textrm{m}_{\textrm{\mistfvariant/}}(G, S) = \textrm{m}_{\textrm{\textrm{\vdkr}}}(G', T)$.

To show that \vdkr is $\APX$-hard in this case, we use an $L$-reduction~\cite{Crescenzi97}. An
 $L$-reduction from optimisation problem $Q$ to an
 optimisation problem $P$ shows that if there exists a
 $(1+\delta)$-approximation algorithm for $P$ then there
 exists a $(1 + \alpha\beta\delta)$-approximation algorithm for $Q$. The reduction that we have described from \mistfvariant/ to \vdkr is an $L$-reduction with $\alpha=\beta=1$ (also called a \emph{strict reduction}~\cite{Crescenzi97}), which shows that \vdkr is $\APX$-hard even when $\Delta(G') = \lceil 5r/3 \rceil - 1$ (shown in Lemma~\ref{lem:krpacking_vdkr_reduction_degree}). To show that \vdkr is $\APX$-hard even when $\Delta(G') \geq \lceil 5r/3 \rceil - 1$, one can add to $G'$ a disconnected star.
% To prove that the reduction is correct we show that the tuple $( f, g, \alpha, \beta )$ meets the five conditions of a correct $L$-reduction~\cite{ausielloetal99}. First, the reduction can be performed in polynomial time. Second, if $G$ is a triangle-free undirected graph with maximum degree $3$ and there exists an independent set in $G$ then by Lemma~\ref{lem:krpacking_reductioniff} there also exists a $K_r$-packing in $G'$. Third, given any undirected graph $G$ and any $K_r$-packing in the constructed graph $G'$, a corresponding independent set $S$ in $G$ can be easily constructed in polynomial time. Fourth, if the size of a maximum $K_r$-packing $S'$ in $G'$ is $B$ then by Lemma~\ref{lem:krpacking_reductioniff} the size of the maximum independent set in $G$ is also $B$. Fifth, also by Lemma~\ref{lem:krpacking_reductioniff}, for any $K_r$-packing $S'$ in the constructed graph $G'$, the difference between the the size of a maximum independent set $S$ in $G$ and the independent set in $G$ corresponding to $S'$ is the same as the difference between the size of a maximum $K_r$-packing in $G'$ and the size of $S'$.
\end{proof}

\subsection{Edge-disjoint packing} 

\subsubsection{Edge-disjoint \texorpdfstring{$K_r$}{Kr}-packing when \texorpdfstring{$r \geq 6$}{r >= 6}}

If $r \geq 6$ and $\Delta = \lceil 5r/3 \rceil - 1$ then it must be that $\Delta < 2r - 2$, so by Observation~\ref{obs:krpacking_edkr_is_also_vdkr}, any edge-disjoint $K_r$-packing is also vertex disjoint. Theorem~\ref{thm:krpacking_edkr_rgeq6_apxhard} then follows, which shows that if $r \geq 6$ then \edkr is $\APX$-hard even when $\Delta \geq \lceil 5r/3 \rceil - 1$. 

% In this section we show that \edkr if $\APX$-hard for any $r \geq 6$ if $\Delta \geq \lceil 5r/3 \rceil - 1$. The proof uses the fact that if $r \geq 6$ then $\lceil 5r/3 \rceil - 1 < 2r - 2$ and thus by Observation~\ref{obs:krpacking_edkr_is_also_vdkr}, any edge-disjoint $K_r$-packing is also vertex-disjoint.

\begin{thm}
\label{thm:krpacking_edkr_rgeq6_apxhard}
If $r \geq 6$ and $\Delta \geq \lceil 5r/3 \rceil - 1$ then \edkr is $\APX$-hard.
\end{thm}
\begin{proof}
Suppose $\Delta = \lceil 5r/3 \rceil - 1$. By definition, any vertex-disjoint $K_r$-packing is also edge disjoint. Since $r \geq 6$ it follows that $\Delta = \lceil 5r/3 \rceil - 1 < 2r - 2$ so by Observation~\ref{obs:krpacking_edkr_is_also_vdkr}, any edge-disjoint $K_r$-packing is also vertex disjoint. We have shown that an edge-disjoint $K_r$-packing of size $B$ exists in $G$ if and only if a vertex-disjoint $K_r$-packing of size $B$ exists in $G$. This fact constitutes an $L$-reduction with $\alpha=\beta=1$ from the restricted case of \vdkr in which $\Delta = \lceil 5r/3 \rceil - 1$. The lemma follows by Theorem~\ref{thm:krpacking_vdkr_apxhard}. As in the proof of Theorem~\ref{thm:krpacking_vdkr_apxhard}, to show that \edkr is $\APX$-hard if $r\geq 6$ even when $\Delta(G) \geq \lceil 5r/3 \rceil - 1$, one can add to $G'$ a disconnected star.
\end{proof}

\subsubsection{Edge-disjoint \texorpdfstring{$K_4$}{K4}-packing} 
\label{sec:krpacking_edkfour}

We now show that \edkfour is $\APX$-hard even when $\Delta=7$. We present an $L$-reduction from a variant of \emph{Maximum Satisfiability}~\cite{ACGKMP99}, inspired by the $L$-reduction of Caprara and Rizzi~\cite{caprara_packing_2002} for \edkthree when $\Delta=5$.

An instance of Maximum Satisfiability is a boolean formula $\phi$ in conjunctive normal form with \emph{clauses} $C=\{ c_1, c_2, \dots, c_{|C|} \}$ and \emph{variables} $X=\{ x_1, x_2, \dots, x_{|X|} \}$. Each clause contains a set of \emph{literals}. Each literal is formed by either a variable or its negation. A \emph{truth assignment} $\mathfrak{f}$ is a function $\mathfrak{f} : X \mapsto \{ \text{true}, \text{false} \}$. A clause is \emph{satisfied} by $\mathfrak{f}$ if any of its literals are true. The goal is to find a truth assignment that satisfies the maximum number of clauses. We denote by \maxtwosatthree/ the special case of Maximum Satisfiability in which each clause contains at most two literals and each variable occurs in at most three clauses. Let $m_i$ be the number of occurrences of variable $x_i$ in $\phi$ for each variable $x_i \in X$. We assume that $2\leq m_i \leq 3$ for each $x_i \in X$, since if any variable occurs in exactly one clause it can be set to the value satisfying that clause. \maxtwosatthree/ is $\APX$-hard~\cite{ACGKMP99}.

Given an instance $\phi$ of \maxtwosatthree/, we construct a graph $G$ such that a truth assignment for $\phi$ exists that satisfies at least $k$ clauses if and only if there exists an edge-disjoint $K_4$-packing of size at least $\sum_{i=1}^{|X|} 3 m_i + k$. As in the case of the reduction presented for \edkthree by Caprara and Rizzi, the reduction here is one of local replacement~\cite{GJ79}. The reduction, shown in Figure~\ref{fig:krpacking_edkfour}, involves the construction and connection of variable and clause gadgets, which is a common technique when reducing from a variant of Maximum Satisfiability. The reduction itself is as follows.

\begin{figure}
    \centering
    \input{figures/edkfour.tikz}
    \caption{The reduction from \maxtwosatthree/ to \edkfour}
    \label{fig:krpacking_edkfour}
\end{figure}

For each variable $x_i$, construct a variable gadget of $10 m_i$ vertices, labelled $R_i = \{ a_i^j,\allowbreak b_i^j,\allowbreak c_i^j,\allowbreak d_i^j,\allowbreak e_i^j,\allowbreak h_i^j,\allowbreak u_i^j,\allowbreak v_i^j,\allowbreak w_i^j,\allowbreak y_i^j \}$ for each $j$ where $1\leq j \leq m_i$. For each $j$ where $1\leq j \leq m_i$, add an edge (if it does not exist already) between each pair of vertices in $\{ a_i^j, b_i^j, u_i^j, v_i^j \}$, $\{ a_i^j, b_i^j, c_i^j, v_i^j \}$, $\{ c_i^j, v_i^j, d_i^j, w_i^j \}$, $\{ d_i^j, w_i^j, e_i^j, y_i^j \}$, $\{ d_i^j, e_i^j, h_i^j, y_i^j \}$, and finally $\{ h_i^j, a_i^{j+1}, y_i^j, u_i^{j+1} \}$ if $j < m_i$ and otherwise $\{ h_i^j, a_i^1, y_i^j, u_i^1 \}$.

We shall refer to $\{ a_i^j, b_i^j, u_i^j, v_i^j \}$, $\{ c_i^j, v_i^j, d_i^j, w_i^j \}$, and $\{ d_i^j, e_i^j, h_i^j, y_i^j \}$ as the \emph{even $K_4$s in $R_i$}, and $\{ a_i^j, b_i^j, c_i^j, v_i^j \}$, $\{ d_i^j, w_i^j, e_i^j, y_i^j \}$, and $\{ h_i^j, a_i^{j+1}, y_i^j, u_i^{j+1} \}$ (and $\{ h_i^j, a_i^1, y_i^j, u_i^1 \}$) as the \emph{odd $K_4$s in $R_i$}. Note that at this point in construction, $\deg_{G}(a_i^j) = \deg_{G}(v_i^j) = \deg_{G}(d_i^j) = \deg_{G}(y_i^j) = 6$, $\deg_{G}(u_i^j) = \deg_{G}(c_i^j) = \deg_{G}(w_i^j) = \deg_{G}(h_i^j) = 5$, and $\deg_{G}(b_i^j) = \deg_{G}(e_i^j) = 4$ for each $j$ where $1\leq j \leq m_i$. 

We shall now construct the clause gadgets. For each clause $c_r$, construct a clause gadget of $5$ vertices labelled $S_r = \{ s_r^1, t_r^1, s_r^2, t_r^2, w_r \}$. Add an edge (if it does not exist already) between each pair of vertices in $\{ s_r^1, t_r^1, s_r^2, w_r \}$ and $\{ s_r^1, s_r^2, t_r^2, w_r \}$. We shall refer to $\{ s_r^1, t_r^1, s_r^2, w_r \}$ and $\{ s_r^1, s_r^2, t_r^2, w_r \}$ as $P_i^r$ and $P_j^r$ supposing the variables of the first and second literals in $c_r$ are $x_i$ and $x_j$. Note that at this point in construction, $\deg_{G}(s_r^1) = \deg_{G}(s_r^2) = \deg_{G}(w_r) = 4$ and $\deg_{G}(t_r^1) = \deg_{G}(t_r^2) = 3$.

We shall now connect the variable and clause gadgets. For each clause $c_r$, suppose $x_i$ is the variable of some literal in $c_r$ where $c_r$ contains the $j\textsuperscript{th}$ occurrence of $x_i$ in $\phi$. If $x_i$ is the first literal in $c_r$ and occurs positively in $c_r$ then identify $b_i^j$ and $s_r^1$, and $c_i^j$ and $t_r^1$. We shall hereafter refer to the first identified vertex as either $b_i^j$ or $s_r^1$ and the second identified vertex as either $c_i^j$ or $t_r^1$. Note that now $\deg_{G}(b_i^j) = \deg_{G}(c_i^j) = 7$. Similarly, if $x_i$ is the first literal in $c_r$ and occurs negatively in $c_r$ then identify $e_i^j$ and $s_r^1$, and $h_i^j$ and $t_r^1$. In this case $\deg_{G}(e_i^j) = \deg_{G}(h_i^j) = 7$. If $x_i$ is the second literal in $c_r$ and occurs positively in $c_r$ then identify $b_i^j$ and $s_r^2$, and $c_i^j$ and $t_r^2$. Similarly, if $x_i$ is the second literal in $c_r$ and occurs negatively in $c_r$ then identify $e_i^j$ and $s_r^2$, and $h_i^j$ and $t_r^2$. This completes the construction of $G$. Observe that $\Delta = 7$.

It is straightforward to show that the reduction can be performed in polynomial time. We now prove that the reduction is correct in the first direction. By construction, no $K_4$ exists in $G$ that contains at least one vertex in a variable gadget and at least one vertex in a clause gadget. Thus, we shall say that some $K_4$ is \emph{in} a variable or clause gadget if it is a strict subset of that gadget.

\begin{lem}
\label{lem:krpacking_kfourpacking_firstdirection}
If a truth assignment $\mathfrak{f}$ for $\phi$ satisfies at least $k$ clauses then an edge-disjoint $K_4$-packing $T$ exists in $G$ where $|T| \geq \sum_{i=1}^{|X|} 3 m_i + k$.
\end{lem}
\begin{proof}
Suppose $\mathfrak{f}$ is a truth assignment for $\phi$ that satisfies at least $k$ clauses. We shall construct an edge-disjoint $K_4$-packing $T$ where $|T| \geq \sum_{i=1}^{|X|} 3 m_i + k$.

For each variable $x_i$, if $\mathfrak{f}(x_i)$ is true then add the set of even $K_4$s in $R_i$ to $T$. Similarly, if $\mathfrak{f}(x_i)$ is false then add the set of odd $K_4$s in $R_i$ to $T$. Now $|T|=\sum_{i=1}^{|X|} 3 m_i$.
For each clause gadget $c_r$ that is satisfied by $\mathfrak{f}$, it must be that there exists some variable $x_i$ where either $\mathfrak{f}(x_i)$ is true and $x_i$ occurs positively in $c_r$ or $\mathfrak{f}(x_i)$ is false and $x_i$ occurs negatively in $c_r$. In either case, add $P_i^r$ to $T$. Now, $T$ contains exactly $\sum_{i=1}^{|X|} 3m_i$ $K_4$s in variable gadgets and at least $k$ $K_4$s in clause gadgets.
It remains to show that $T$ is edge disjoint. By the construction of $G$, any two $K_4$s in $T$ in the same variable gadget are edge disjoint. Consider an arbitrary $P_r^i$ in some clause gadget $c_r$ that belongs to $T$. It must be that either $\mathfrak{f}(x_i)$ is true and $x_i$ occurs positively in $c_r$ or $\mathfrak{f}(x_i)$ is false and $x_i$ occurs negatively in $c_r$. In the former case, $T$ contains the set of even $K_4$s in $R_i$ so since $P_i^r \cap R_i = \{ b_i^j, c_i^j \}$ where $1\leq j\leq 3$ it follows that $T$ is edge disjoint. In the latter case, $T$ contains the set of odd $K_4$s in $R_i$ so since $P_i^r \cap R_i = \{ e_i^j, h_i^j \}$ where $1\leq j\leq 3$ it also follows that $T$ is edge disjoint.
\end{proof}

We now prove that the reduction is correct in the second direction. We say that some edge-disjoint $K_4$-packing $T$ in $G$ is \emph{canonical} if for any variable gadget $R_i$, $T$ contains either the set of even $K_4$s in $R_i$ or the set of odd $K_4$s in $R_i$. By the construction of $G$, no edge-disjoint $K_4$-packing can contain all even $K_4$s and all odd $K_4$s.

We first show that for any variable gadget $R_i$ and edge-disjoint $K_r$-packing $T$, if $T$ contains neither all even $K_4$s in $R_i$ nor all odd $K_4$s in $R_i$ then the number of $K_4$s in $T$ is at most $3 m_i - 1$.

\begin{prop}
\label{prop:krpacking_kfour_evenoddareonlymaximum}
Suppose $T$ is an arbitrary edge-disjoint $K_4$-packing in $G$. For any variable gadget $R_i$, if $T$ contains neither all even $K_4$s in $R_i$ nor all odd $K_4$s in $R_i$ then the number of $K_4$s in $T$ is at most $3 m_i - 1$.
\end{prop}
\begin{proof}
By the construction of $G$, each even $K_4$ in $R_i$ intersects exactly two odd $K_4$s in $R_i$ by at least two vertices and each odd $K_4$ in $R_i$ intersects exactly two even $K_4$s in $R_i$ by at least two vertices. 

It follows that the $K_4$-edge intersection graph ${\mathcal{K}'}_r^G$ contains a cycle of $6 m_i$ vertices corresponding to the $6 m_i$ $K_4$s in $R_i$. It then follows that any edge-disjoint $K_4$-packing that contains $3 m_i$ $K_4$s in $R_i$ corresponds to an independent set of size $3 m_i$ in ${\mathcal{K}'}_r^G$, and thus is either the set of even $K_4$s in $R_i$ or the set of odd $K_4$s in $R_i$. Since $T$ contains neither all even $K_4$s in $R_i$ nor all odd $K_4$s in $R_i$ it follows that $|T| < 3m_i$.
\end{proof}

We can now prove that for any edge-disjoint $K_4$-packing in $G$ that is not canonical, there exists a canonical edge-disjoint $K_4$-packing in $G$ of at least the same cardinality.

\begin{lem}
If $T$ is an edge-disjoint $K_4$-packing then there exists a canonical edge-disjoint $K_4$-packing $T'$ where $|T'| \geq |T|$.
\label{lem:krpacking_four_canonical}
\end{lem}
\begin{proof}
If $T$ is already canonical then let $T'=T$. Otherwise, by the definition of canonical, there must exist at least one variable gadget $i$ such that $T$ contains neither all even $K_4$s in $R_i$ nor all odd $K_4$s in $R_i$. For any such $i$ where $1\leq i \leq |X|$, we show how to modify $T$ to ensure that it either contains the set of even $K_4$s in $R_i$ or the set of odd $K_4$s in $R_i$ and the cardinality of $T$ does not decrease. It follows that there exists a canonical edge-disjoint $K_4$-packing $T'$ where $|T'| \geq |T|$. 

Note that by Proposition~\ref{prop:krpacking_kfour_evenoddareonlymaximum}, the number of $K_4$s in $R_i$ in $T$ is at most $3 m_i - 1$. 

Suppose the variable $x_i$ corresponding to $R_i$ occurs in clauses $c_{r_1}, c_{r_2}, \dots, c_{r_{m_i}}$, corresponding to the sets $P_i^{r_1}, P_i^{r_2}, \dots, P_i^{r_{m_i}}$. It must be that either at most one $K_4$ in $\{ P_i^{r_1}, P_i^{r_2}, \dots, P_i^{r_{m_i}} \}$ exists in $T$ where the corresponding occurrence of $x_i$ is positive; or at most one $K_4$ in $\{ P_i^{r_1}, P_i^{r_2}, \dots, P_i^{r_{m_i}} \}$ exists in $T$ where the corresponding occurrence of $x_i$ is negative. Suppose the former case is true. Remove the $K_4$ in $\{ P_i^{r_1}, P_i^{r_2}, \dots, P_i^{r_{m_i}} \}$ in $T$ where the corresponding occurrence of $x_i$ is positive. Next, remove any even $K_4$s in $R_i$ in $T$ and add the set of odd $K_4$s in $R_i$ not already in $T$. The number of $K_4$s in $R_i$ in $T$ is now $3 m_i$ so since at most one $K_4$ was removed, which was not in $R_i$, it follows that the cardinality of $T$ has not decreased. To see that $T$ is still edge-disjoint, observe that any $K_4$ in $\{ P_i^{r_1}, P_i^{r_2}, \dots, P_i^{r_{m_i}} \}$ in $T$ intersects any odd $K_4$ in $R_i$ by at most one vertex. The construction and proof in the latter case are symmetric.
\end{proof}

\begin{lem}
\label{lem:krpacking_kfourpacking_seconddirection}
If $T$ is an edge-disjoint $K_4$-packing where $|T| = \sum_{i=1}^{|X|} 3 m_i + k$ for some integer $k\geq 1$ then exists a truth assignment $\mathfrak{f}$ that satisfies at least $k$ clauses.
\end{lem}
\begin{proof}
Assume by Lemma~\ref{lem:krpacking_four_canonical} that $T$ is canonical. It follows that $T$ contains exactly $\sum_{i=1}^{|X|} 3 m_i$ $K_4$s in variable gadgets and at least $k$ $K_4$s in clause gadgets. For each variable $x_i$, set $\mathfrak{f}(x_i)$ to be true if $T$ contains all even $K_4$s in $R_i$ and false otherwise. Now consider each clause gadget $c_r$ where $S_r$ contains some $K_4$ in $T$, denoted $P_i^r$. Suppose $x_i$ occurs positively in $c_r$. It follows that $P_i^r$ contains $b_i^j, c_i^j$ for some $j$ where $1\leq j\leq 3$. Since $T$ is canonical and edge-disjoint it follows that $T$ contains the set of even $K_4$s in $R_i$. By the construction of $\mathfrak{f}$ it follows that $\mathfrak{f}(x_i)$ is true and thus $c_r$ is satisfied. The proof for when $x_i$ occurs negatively in $c_r$ is symmetric. It follows that at least $k$ clauses are satisfied by $\mathfrak{f}$.
\end{proof}

\begin{lem}
\label{lem:krpacking_edkr_req4_apxhard}
If $r=4$ and $\Delta=7$ then \edkr is $\APX$-hard.
\end{lem}
\begin{proof}
We now show that the $L$-reduction we have described from \maxtwosatthree/ (which is $\APX$-hard~\cite{ACGKMP99}) to \edkfour when $\Delta = 7$ is valid. For compactness, we abbreviate \maxtwosatthree/ when appearing in a subscript to \maxtwosatthreeshort/.

An $L$-reduction is characterised by a pair $(f, g)$ of functions that can be computed in polynomial time. Here, $f$ is the reduction described at the start of this section (Section~\ref{sec:krpacking_edkfour}) in which an instance $G$ of \edkfour is constructed from an arbitrary instance $\phi$ of \maxtwosatthree/. It is straightforward to show that $f$ can be computed in polynomial time.

The function $g$ is described by Lemma~\ref{lem:krpacking_kfourpacking_seconddirection}. For any instance $\phi$ of \maxtwosatthree/ and edge-disjoint $K_4$-packing in $f(\phi)$, $g$ computes a truth assignment $\mathfrak{f}$ for $\phi$. It is also straightforward to show that $g$ can be computed in polynomial time.

To show that $f$ and $g$ constitute a valid $L$-reduction, we must show that there exists fixed constants $\alpha$ and $\beta$ such that for any instance $\phi$ of \maxtwosatthree/,
\begin{align*}
    \textrm{opt}_{\textrm{\edkfour}}(f(\phi)) \leq \alpha \textrm{opt}_{\textrm{\maxtwosatthreeshort/}}(\phi)
\end{align*}
and that for any instance $\phi$ and any edge-disjoint $K_4$-packing $T$ in $f(\phi)$,
\begin{align*}
    \textrm{opt}_{\textrm{\maxtwosatthreeshort/}}(\phi) - \textrm{m}_{\textrm{\maxtwosatthreeshort/}}(\phi, g(\phi, T)) \leq \beta(\textrm{opt}_{\textrm{\edkfour}}(f(\phi)) - \textrm{m}_{\textrm{\edkfour}}(f(\phi), T))\enspace.
\end{align*}
We shall now demonstrate the existence of some such $\alpha$ and $\beta$. Recall that in the instance of \maxtwosatthree/, $X$ is the set of variables, $C$ is the set of clauses, and $m_i$ is the number of occurrences of each variable $x_i$. Note that by the definition of \maxtwosatthree/, $\sum_{i=1}^{|X|} m_i$ is the total number of literals, which must be at most $2|C|$. Note also that for any instance $\phi$ of \maxtwosatthree/, it must be that $\textrm{opt}_{\textrm{\maxtwosatthreeshort/}}(\phi) \geq |C|/2$. This is because a truth assignment satisfying $|C|/2$ clauses can be found using a greedy algorithm that in each step assigns a truth value to a variable occurring in the maximum number of clauses~\cite{approximationvazirani}. We can now show that
\begin{align*}
    \textrm{opt}_{\textrm{\edkfour}}(f(\phi)) &\leq \sum\limits_{i=1}^{|X|} 3 m_i + \textrm{opt}_{\textrm{\maxtwosatthreeshort/}}(\phi) && \mbox{by Lemma~\ref{lem:krpacking_kfourpacking_seconddirection}}\\
    &= 3 \sum\limits_{i=1}^{|X|} m_i + \textrm{opt}_{\textrm{\maxtwosatthreeshort/}}(\phi)\\
    &\leq 6|C| + \textrm{opt}_{\textrm{\maxtwosatthreeshort/}}(\phi) && \mbox{since $2|C| \geq \sum_{i=1}^{|X|} m_i$}\\
    &\leq 13\textrm{opt}_{\textrm{\maxtwosatthreeshort/}}(\phi)  && \mbox{since $\textrm{opt}_{\textrm{\maxtwosatthreeshort/}}(\phi) \geq |C|/2$}
\end{align*}
so $\alpha = 13$. We can also show that for any instance $\phi$ and any edge-disjoint $K_4$-packing $T$ in $f(\phi)$, 
\begin{align*}
    \textrm{opt}_{\textrm{\maxtwosatthreeshort/}}(\phi) - \textrm{m}_{\textrm{\maxtwosatthreeshort/}}(\phi, g(\phi, T)) &\leq \textrm{opt}_{\textrm{\maxtwosatthreeshort/}}(\phi) - \left(|T| - \sum\limits_{i=1}^{|X|} 3 m_i \right) && \mbox{by Lemma~\ref{lem:krpacking_kfourpacking_seconddirection}}\\
    &= \sum\limits_{i=1}^{|X|} 3 m_i + \textrm{opt}_{\textrm{\maxtwosatthreeshort/}}(\phi) - |T|\\
    &\leq \textrm{opt}_{\textrm{\edkfour}}(f(\phi)) - |T| && \mbox{by Lemma~\ref{lem:krpacking_kfourpacking_firstdirection}}\\
    &= \textrm{opt}_{\textrm{\edkfour}}(f(\phi)) - \textrm{m}_{\textrm{\edkfour}}(f(\phi), T) && \mbox{since $\textrm{m}_{\textrm{\edkfour}}(f(\phi), T) = |T|$}
\end{align*}
which shows that $\beta = 1$.
\end{proof}

\subsubsection{Edge-disjoint \texorpdfstring{$K_5$}{K5}-packing} 

We now show that \edkfive is $\APX$-hard even when $\Delta=9$. We present an $L$-reduction that follows the same pattern as the one shown in Section~\ref{sec:krpacking_edkfour}. This reduction, shown in Figure~\ref{fig:krpacking_edkfive}, is as follows.
\begin{figure}[t]
    \centering
    \input{figures/edkfive.tikz}
    \caption{The reduction from \maxtwosatthree/ to \edkfive}
    \label{fig:krpacking_edkfive}
\end{figure}
For each variable $x_i$, construct a variable gadget of $8 m_i$ vertices, labelled $R_i = \{ a_i^j, b_i^j, c_i^j, d_i^j, e_i^j, h_i^j, u_i^j, v_i^j \}$ for each $j$ where $1\leq j \leq m_i$. For each $j$ where $1\leq j \leq m_i$, add an edge (if it does not exist already) between each pair of vertices in $\{ a_i^j, b_i^j, e_i^j, u_i^j, v_i^j \}$, $\{ b_i^j, c_i^j, e_i^j, h_i^j, v_i^j \}$, and finally $\{ c_i^j, d_i^j, h_i^j, v_i^j, u_i^{j+1} \}$ and $\{ d_i^j, a_i^{j+1}, h_i^j, e_i^{j+1}, u_i^{j+1} \}$ if $j < m_i$, otherwise $\{ c_i^j, d_i^j, h_i^j, v_i^j, u_i^1 \}$ and $\{ d_i^j, a_i^1, h_i^j, e_i^1, u_i^1 \}$. We shall refer to $\{ a_i^j, b_i^j, e_i^j, u_i^j, v_i^j \}$ and $\{ c_i^j, d_i^j, h_i^j, v_i^j, u_i^{j+1} \}$ (and $\{ c_i^j, d_i^j, h_i^j, v_i^j, u_i^1 \}$) as \emph{odd $K_5$s in $R_i$}, and $\{ b_i^j, c_i^j, e_i^j, h_i^j, v_i^j \}$ and $\{ d_i^j, a_i^{j+1}, h_i^j, e_i^{j+1}, u_i^{j+1} \}$ (and $\{ d_i^j, a_i^1, h_i^j, e_i^1, u_i^1 \}$) as \emph{even $K_5$s in $R_i$}. At this point $\deg_{G}(a_i^j) = \deg_{G}(b_i^j) = \deg_{G}(c_i^j) = \deg_{G}(d_i^j) = 6$ and $\deg_{G}(e_i^j) = \deg_{G}(h_i^j) = \deg_{G}(u_i^j) = \deg_{G}(v_i^j) = 8$ for any $j$ where $1\leq j \leq m_i$. 

% We begin by constructing the variable gadgets. For each variable $x_i$, construct $10m_i$ vertices labelled $a_i^1, b_i^1, c_i^1, d_i^1, e_i^1, h_i^1, u_i^1, v_i^1, a_i^2, b_i^2, \dots, u_i^{m_i}, v_i^{m_i}$. We shall refer to these vertices as the \emph{$i\textsuperscript{th}$ variable gadget}. For each $j$ where $1\leq j \leq m_i$, add an edge (if it does not exist already) between each pair of vertices in $\{ a_i^j, b_i^j, e_i^j, u_i^j, v_i^j \}$, $\{ b_i^j, c_i^j, e_i^j, h_i^j, v_i^j \}$, and finally $\{ c_i^j, d_i^j, h_i^j, v_i^j, u_i^{j+1} \}$ and $\{ d_i^j, a_i^{j+1}, h_i^j, e_i^{j+1}, u_i^{j+1} \}$ if $j < m_i$, otherwise $\{ c_i^j, d_i^j, h_i^j, v_i^j, u_i^1 \}$ and $\{ d_i^j, a_i^1, h_i^j, e_i^1, u_i^1 \}$. We shall refer to $\{ a_i^j, b_i^j, e_i^j, u_i^j, v_i^j \}$ and $\{ c_i^j, d_i^j, h_i^j, v_i^j, u_i^{j+1} \}$ (and $\{ c_i^j, d_i^j, h_i^j, v_i^j, u_i^1 \}$) as \emph{even} $K_5$s, and $\{ b_i^j, c_i^j, e_i^j, h_i^j, v_i^j \}$ and $\{ d_i^j, a_i^{j+1}, h_i^j, e_i^{j+1}, u_i^{j+1} \}$ (and $\{ d_i^j, a_i^1, h_i^j, e_i^1, u_i^1 \}$) as \emph{odd} $K_5$s. Note that at this point of construction, $\deg_{G}(a_i^j) = \deg_{G}(b_i^j) = \deg_{G}(c_i^j) = \deg_{G}(d_i^j) = 6$ and $\deg_{G}(e_i^j) = \deg_{G}(h_i^j) = \deg_{G}(u_i^j) = \deg_{G}(v_i^j) = 8$ for any $j$ where $1\leq j \leq m_i$. 

We shall now construct the clause gadgets. For each clause $c_r$, construct a clause gadget of $7$ vertices labelled $S_r = \{ s_r^1, t_r^1, s_r^2, t_r^2, w_r^1, w_r^2, w_r^3, w_r^4 \}$. Add an edge (if it does not exist already) between each pair of vertices in $\{ s_r^1, t_r^1, w_r^1, w_r^2, w_r^3 \}$ and $\{ s_r^2, t_r^2, w_r^2, w_r^3, w_r^4 \}$. Label $\{ s_r^1, t_r^1, w_r^1, w_r^2, w_r^3 \}$ and $\{ s_r^2, t_r^2, w_r^2, w_r^3, w_r^4 \}$ as $P_i^r$ and $P_j^r$, where the variables of the literals in $c_r$ are $x_i$ and $x_j$.

% We shall now construct the clause gadgets. For each clause $c_r$, construct $7$ vertices labelled $s_r^1, t_r^1, s_r^2, t_r^2, w_r^1, w_r^2, w_r^3, w_r^4$. We shall refer to these vertices as the \emph{$r\textsuperscript{th}$ clause gadget}. Add an edge (if it does not exist already) between each pair of vertices in $\{ s_r^1, t_r^1, w_r^1, w_r^2, w_r^3 \}$ and $\{ s_r^2, t_r^2, w_r^2, w_r^3, w_r^4 \}$. We shall refer to $\{ s_r^1, t_r^1, w_r^1, w_r^2, w_r^3 \}$ and $\{ s_r^2, t_r^2, w_r^2, w_r^3, w_r^4 \}$ as $P_i^r$ and $P_j^r$ where the variables of the literals in $c_r$ are $x_i$ and $x_j$.

The connection of variable and clause gadgets follows the same pattern as for \edkfour. For each clause $c_r$, suppose $x_i$ is the variable of some literal in $c_r$ where $c_r$ contains the $j\textsuperscript{th}$ occurrence of $x_i$ in $\phi$. If $x_i$ is the first literal in $c_r$ and occurs positively in $c_r$ then identify $a_i^j$ and $s_r^1$, and $b_i^j$ and $t_r^1$. Now $\deg_{G}(a_i^j) = \deg_{G}(b_i^j) = 9$. Similarly, if $x_i$ is the first literal in $c_r$ and occurs negatively in $c_r$ then identify $b_i^j$ and $s_r^1$, and $c_i^j$ and $t_r^1$. If $x_i$ is the second literal in $c_r$ and occurs positively in $c_r$ then identify $a_i^j$ and $s_r^2$, and $b_i^j$ and $t_r^2$. If $x_i$ is the second literal in $c_r$ and occurs negatively in $c_r$ then identify $b_i^j$ and $s_r^2$, and $c_i^j$ and $t_r^2$. Now $\Delta = 9$.

% We shall now connect the variable and clause gadgets. For each clause $c_r$, suppose $x_i$ is the variable of the some literal in $c_r$ where $c_r$ contains the $j\textsuperscript{th}$ occurrence of $x_i$ in $\phi$. If $x_i$ is the first literal in $c_r$ and occurs positively in $c_r$ then identify $a_i^j$ and $s_r^1$, and $b_i^j$ and $t_r^1$. We shall hereafter refer to the first identified vertex as either $a_i^j$ or $s_r^1$ and the second identified vertex as either $b_i^j$ or $t_r^1$. Note that now $\deg_{G}(a_i^j) = \deg_{G}(b_i^j) = 9$. Similarly, if $x_i$ is the first literal in $c_r$ and occurs negatively in $c_r$ then identify $b_i^j$ and $s_r^1$, and $c_i^j$ and $t_r^1$. If $x_i$ is the second literal in $c_r$ and occurs positively in $c_r$ then identify $a_i^j$ and $s_r^2$, and $b_i^j$ and $t_r^2$. If $x_i$ is the second literal in $c_r$ and occurs negatively in $c_r$ then identify $b_i^j$ and $s_r^2$, and $c_i^j$ and $t_r^2$. This completes the construction of $G$. Observe that $\Delta = 9$.

As before, the reduction can be performed in polynomial time.  We now prove correctness in the first direction.

\begin{lem}
\label{lem:krpacking_kfivepacking_firstdirection}
If a truth assignment $\mathfrak{f}$ for $\phi$ satisfies at least $k$ clauses then an edge-disjoint $K_5$-packing $T$ exists in $G$ where $|T| \geq \sum_{i=1}^{|X|} 2 m_i + k$.
\end{lem}
\begin{proof}
Suppose $\mathfrak{f}$ is a truth assignment for $\phi$ that satisfies at least $k$ clauses. We shall construct an edge-disjoint $K_5$-packing $T$ where $|T| \geq \sum_{i=1}^{|X|} 2 m_i + k$.
For each variable $x_i$, add to $T$ the set of even $K_5$s in $R_i$ if $\mathfrak{f}(x_i)$ is true and otherwise the set of odd $K_5$s in $R_i$. Now $|T|=\sum_{i=1}^{|X|} 3 m_i$.
For each clause $c_r$ satisfied by $\mathfrak{f}$, it must be that there exists some variable $x_i$ where $\mathfrak{f}(x_i)$ is true and $x_i$ occurs positively in $c_r$, or there exists some variable $x_i$ where $\mathfrak{f}(x_i)$ is false and $x_i$ occurs negatively in $c_r$. As before, in either case add $P_i^r$ to $T$. Now $|T| = \sum_{i=1}^{|X|} 2m_i + k$. The proof that $T$ is edge disjoint is analogous to the proof in Lemma~\ref{lem:krpacking_kfourpacking_firstdirection}.
\end{proof}

% We now prove that the reduction is correct in the second direction. In the remainder of this section, assume that $T$ is an edge-disjoint $K_r$-packing where $|T|=\sum_{i=1}^{|X|} 2m_i + k$ for some integer $k\geq 1$. We shall eventually construct a truth assignment $\mathfrak{f}$ that satisfies at least $|T|=\sum_{i=1}^{|X|} 2m_i + k$ clauses.

We now prove the second direction. Like before, we say that some edge-disjoint $K_5$-packing $T$ in $G$ is \emph{canonical} if for any $R_i$, $T$ contains either the set of even $K_5$s in $R_i$ or the set of odd $K_5$s in $R_i$.

\begin{lem}
If $T$ is an edge-disjoint $K_5$-packing then there exists a canonical edge-disjoint $K_5$-packing $T'$ where $|T'| \geq |T|$.
\label{lem:krpacking_five_canonical}
\end{lem}
\begin{proof}
The proof is analogous to the proof of Lemma~\ref{lem:krpacking_four_canonical}. Here we describe the modification of a single variable gadget $R_i$ where $T$ contains neither all even $K_5$s nor all odd $K_5$s in $R_i$. It must be that the number of $K_5$s in $R_i$ in $T$ is at most $2 m_i - 1$.

% By the definition of canonical, there exists at least one variable gadget $i$ such that $T$ contains neither all even $K_5$s in the $i\textsuperscript{th}$ variable gadget nor all odd $K_5$s in the $i\textsuperscript{th}$ variable gadget. For any such $i$, we show how to modify $T$ to construct $T'$ such that $|T'|\geq |T|$ and $T'$ contains either all even $K_5$s in $i\textsuperscript{th}$ variable gadget or all odd $K_5$s in the $i\textsuperscript{th}$ variable gadget. It follows from this that there exists some canonical edge-disjoint $K_r$-packing $T'$ where $|T'| \geq |T|$. 

% In the remainder of this proof we shall consider only one variable gadget, namely the $i\textsuperscript{th}$. It must be that at most $2m_i - 1$ $K_5$s that are subsets of this variable gadget belong to $T$.

Suppose $x_i$ occurs in clauses $c_{r_1}, c_{r_2}, \dots, c_{r_{m_i}}$, corresponding to the sets $P_i^{r_1}, P_i^{r_2}, \dots, P_i^{r_{m_i}}$. It must be that either at most one $K_5$ in $\{ P_i^{r_1}, P_i^{r_2}, \dots, P_i^{r_{m_i}} \}$ exists in $T$ where the corresponding occurrence of $x_i$ is positive, or at most one $K_5$ in $\{ P_i^{r_1}, P_i^{r_2}, \dots, P_i^{r_{m_i}} \}$ exists in $T$ where the corresponding occurrence of $x_i$ is negative. In the former case, remove the $K_5$ in $\{ P_i^{r_1}, P_i^{r_2}, \dots, P_i^{r_{m_i}} \}$ where the corresponding occurrence of $x_i$ is positive as well as any even $K_5$s in $R_i$ in $T$, then add the set of odd $K_5$s not already in $T$. The number of $K_5$s in $R_i$ is now $2 m_i$ so since at most one $K_5$ was removed, which was not in $R_i$, it follows that the cardinality of $T$ has not decreased. To see that $T$ is still edge-disjoint, observe that any $K_5$ in $\{ P_i^{r_1}, P_i^{r_2}, \dots, P_i^{r_{m_i}} \}$ in $T$ intersects any odd $K_5$ by at most one vertex. The construction and proof in the latter case is symmetric.
\end{proof}

\begin{lem}
\label{lem:krpacking_kfivepacking_seconddirection}
If $T$ is an edge-disjoint $K_5$-packing where $|T| = \sum_{i=1}^{|X|} 2 m_i + k$ for some integer $k\geq 1$ then exists a truth assignment $\mathfrak{f}$ that satisfies at least $k$ clauses.
\end{lem}
\begin{proof}
Assume by Lemma~\ref{lem:krpacking_five_canonical} that $T$ is canonical. It follows that $T$ contains exactly $\sum_{i=1}^{|X|} 2 m_i$ $K_5$s in variable gadgets and at least $k$ $K_5$s in clause gadgets. For each variable $x_i$, set $\mathfrak{f}(x_i)$ to be true if $T$ contains all even $K_5$s in $R_i$ and false otherwise. Now consider each clause gadget $c_r$ where $S_r$ contains some $K_5$ in $T$, which we label $P_i^r$. Suppose $x_i$ occurs positively in $c_r$. It follows that $P_i^r$ contains $a_i^j, b_i^j$ for some $j$ where $1\leq j\leq 3$. Since $T$ is edge disjoint it follows that $T$ contains the even $K_5$s in $R_i$. By the construction of $\mathfrak{f}$ it follows that $\mathfrak{f}(x_i)$ is true and thus that $c_r$ is satisfied. The proof when $x_i$ occurs negatively in $c_r$ is symmetric. It follows thus that at least $k$ clauses are satisfied by $\mathfrak{f}$.
\end{proof}

\begin{lem}
\label{lem:krpacking_edkr_req5_apxhard}
If $r=5$ and $\Delta=9$ then \edkr is $\APX$-hard.
\end{lem}
\begin{proof}
The reduction described runs in polynomial time, and Lemma~\ref{lem:krpacking_kfivepacking_seconddirection} shows how to construct a truth assignment $\mathfrak{f}$ that satisfies $k$ clauses given an edge-disjoint $K_5$-packing of cardinality $\sum_{i=1}^{|X|} 3 m_i + k$ where $k \geq 1$. By Lemmas~\ref{lem:krpacking_kfivepacking_firstdirection} and~\ref{lem:krpacking_kfivepacking_seconddirection}, in the reduction a truth assignment $\mathfrak{f}$ for $\phi$ exists that satisfies at least $k$ clauses if and only if there exists an edge-disjoint $K_5$-packing of size at least $\sum_{i=1}^{|X|} 3 m_i + k$. This reduction is thus an $L$-reduction with $\alpha=9$ and $\beta=1$.
\end{proof}

We now combine Lemmas~\ref{lem:krpacking_edkr_req4_apxhard} and \ref{lem:krpacking_edkr_req5_apxhard} with the existing result of Caprara and Rizzi~\cite{caprara_packing_2002} in Theorem~\ref{thm:krpacking_edkr345apxhard}.

\begin{thm}
\label{thm:krpacking_edkr345apxhard}
If $3 \leq r \leq 5$ and $\Delta > 2r - 2$ then \edkr is $\APX$-hard.
\end{thm}
\begin{proof}
Caprara and Rizzi~\cite{caprara_packing_2002} prove the case when $r = 3$ and $\Delta = 5$. In Lemma~\ref{lem:krpacking_edkr_req4_apxhard} we prove the case when $r=4$ and $\Delta=7$. In Lemma~\ref{lem:krpacking_edkr_req5_apxhard} we prove the case when $r=5$ and $\Delta=9$.
\end{proof}

% \subsubsection{Conclusion} 

% We now combine Lemmas~\ref{lem:krpacking_edkr_req4_apxhard}, \ref{lem:krpacking_edkr_req5_apxhard}, and \ref{lem:krpacking_edkr_rgeq6_apxhard} in Theorem~\ref{thm:krpacking_edkr_conclusion_apxhard}.

% \begin{thm}
% \label{thm:krpacking_edkr_conclusion_apxhard}
% For any simple undirected graph $G$, If either $r \leq 5$ and $\Delta > 2r - 2$, or $r > 5$ and $\Delta \geq \lceil 5r/3 \rceil - 1$, then \edkr is $\APX$-hard.
% \end{thm}
% \begin{proof}
% Lemma~\ref{lem:krpacking_edkr_rgeq6_apxhard} shows the case when $r\geq 6$ and $\Delta \geq \lceil 5r/3 \rceil - 1$.
% \end{proof}

%Construct $r|V| - r/3|E|$ vertices in $V'$ labelled $w_1, w_2, \dots, w_{|V'|}$. 

%% file: figures/edkfour.tikz
\newcommand\Kfourdraw[4]{%
    \draw (#1) -- (#2) -- (#3) -- (#4) -- (#1);
    \draw (#1) -- (#3);
    \draw (#2) -- (#4);
}

\tikzset{gradientpath/.style n args={3}{
    postaction={
    decorate,
    decoration={
    markings,
    mark=between positions 0 and \pgfdecoratedpathlength step 0.2pt with {
    \pgfmathsetmacro\myval{multiply(
        divide(
        \pgfkeysvalueof{/pgf/decoration/mark info/distance from start}, \pgfdecoratedpathlength
        ),
        100
    )};
    \pgfsetfillcolor{#3!\myval!#2};
    \pgfpathcircle{\pgfpointorigin}{#1};
    \pgfusepath{fill};}
}}}}

\begin{tikzpicture}

% \filldraw[color=red, fill=none](0.0, 0.0) circle (\innerradius);
% \filldraw[color=red, fill=none](0.0, 0.0) circle (\outerradius);

\begin{scope}[shift={(-4.0, 0.0)}, every node/.style={thick, circle, draw, minimum size=2.4mm, fill=white}]

\def\innerradius{4.4}
\def\outerradius{6.4}

\node[draw=none] (hijminus1) at (105:{\outerradius}) {};
\node[draw=none, label={[label distance=0.4cm]90:$a_i^j$}] (aij) at (90:{\outerradius}) {};
\node[label={[label distance=0.4cm]80:$b_i^j$}] (bij) at (75:{\outerradius}) {};
\node[draw=none, label={[label distance=0.4cm]70:$c_i^j$}] (cij) at (60:{\outerradius}) {};
\node[draw=none, label={[label distance=0.4cm]60:$d_i^j$}] (dij) at (45:{\outerradius}) {};
\node[label={[label distance=0.4cm]50:$e_i^j$}] (eij) at (30:{\outerradius}) {};
\node[draw=none, label={[label distance=0.4cm]40:$h_i^j$}] (hij) at (15:{\outerradius}) {};
\node[draw=none] (aij1) at (0:{\outerradius}) {};

\node[draw=none] (yijminus1) at (112.5:{\innerradius}) {};
\node[label={[label distance=0.4cm]270:$u_i^j$}] (uij) at (90:{\innerradius}) {};
\node[label={[label distance=0.4cm]247.5:$v_i^j$}] (vij) at (67.5:{\innerradius}) {};
\node[label={[label distance=0.4cm]225:$w_i^j$}] (wij) at (45:{\innerradius}) {};
\node[label={[label distance=0.4cm]202.5:$y_i^j$}] (yij) at (22.5:{\innerradius}) {};
\node[draw=none] (uij1) at (0:{\innerradius}) {};

\path[gradientpath={0.2pt}{black}{white}] (aij) -- (hijminus1);
\path[gradientpath={0.2pt}{black}{white}] (aij) -- (yijminus1);
\path[gradientpath={0.2pt}{black}{white}] (uij) -- (yijminus1);
\path[gradientpath={0.2pt}{black}{white}] (uij) -- (hijminus1);
% \draw (yijminus1) -- (hijminus1);
\Kfourdraw{aij}{bij}{uij}{vij}
\Kfourdraw{cij}{vij}{dij}{wij}
\draw (bij) -- (cij);
\draw (dij) -- (eij);
\draw (wij) -- (yij);
\draw (dij) -- (yij);
\draw (wij) -- (eij);
\draw (eij) -- (yij);
\draw (eij) -- (hij);
\draw (yij) -- (hij);

\path[gradientpath={0.2pt}{black}{white}] (hij) -- (aij1);
\path[gradientpath={0.2pt}{black}{white}] (yij) -- (uij1);
\path[gradientpath={0.2pt}{black}{white}] (hij) -- (uij1);
\path[gradientpath={0.2pt}{black}{white}] (yij) -- (aij1);

\draw plot [smooth] coordinates {(aij) ($ (bij) !.4! (vij) $) (cij)};
\draw plot [smooth] coordinates {(dij) ($ (eij) !.4! (yij) $) (hij)};

% redraw a,c,d,h
\node (aijfake) at (aij) {};
\node (cijfake) at (cij) {};
\node (dijfake) at (dij) {};
\node (hijfake) at (hij) {};

\begin{scope}[shift={(22.5:3.8)}]
\begin{scope}[shift={(-67.5:2.6)}]
\node[label={[label distance=0.4cm]-67.5:$w_r$}] (wr) at (30:{\outerradius}) {};
\end{scope}
\end{scope}

\begin{scope}[shift={(22.5:2.6)}]
\node[label={[label distance=0.4cm]112.5:$s_r^1$}] (sr1) at (30:{\outerradius}) {};
\node[label={[label distance=0.4cm]-67.5:$t_r^1$}] (tr1) at (15:{\outerradius}) {};
\end{scope}

\begin{scope}[shift={(22.5:5.0)}]
\node[label={[label distance=0.4cm]112.5:$s_r^2$}] (sr2) at (30:{\outerradius}) {};
\node[label={[label distance=0.4cm]-67.5:$t_r^2$}] (tr2) at (15:{\outerradius}) {};
\end{scope}

\draw (sr1) -- (tr1) -- (wr) -- (tr2) -- (sr2) -- cycle;
\draw (tr1) -- (sr2);
\draw (tr2) -- (sr1);
\draw (wr) -- (sr1);
\draw (wr) -- (sr2);
\draw (sr1) -- (sr2);
\end{scope}

% clause gadget

% \begin{scope}[shift={(4.0, 0.0)}]
% \draw[help lines] (0,0) grid (4,4);
% \end{scope}

% draw the gradient, based on https://tex.stackexchange.com/q/606045
% \begin{scope}
% \draw[red] (0,0) ++(90:\outerradius) arc (90:105:\outerradius);
% \def\startangle{90}
% \def\changeangle{22.5}
% \def\inter{1}
% \begin{scope}
%     \foreach \i in {0,\inter,...,\changeangle}
%         {
%         \pgfmathsetmacro\ix{\i+\startangle}
%         \pgfmathsetmacro\colorvalue{\i/\changeangle}
%         \definecolor{slicecolor}{rgb}{\colorvalue,\colorvalue,\colorvalue}
%         \pgfmathsetmacro\jx{\i+\inter+\startangle}
%         \filldraw[thin,red,fill opacity=\colorvalue, draw=none] (0:0) -- ((\ix:\outerradius) arc (\ix:\jx:\outerradius) -- (0:0) -- cycle;
%         }
% \end{scope}
% \end{scope}

\end{tikzpicture}

%% file: figures/edkfive.tikz
\tikzset{
  laser beam action/.style={
    line width=\pgflinewidth+1.0pt,draw opacity=.12,draw=#1,
  },
  laser beam recurs/.code 2 args={%
    \pgfmathtruncatemacro{\level}{#1-1}%
    \ifthenelse{\equal{\level}{0}}%
    {\tikzset{preaction={laser beam action=#2}}}%
    {\tikzset{preaction={laser beam action=#2,laser beam recurs={\level}{#2}}}}
  },
  laser beam/.style={preaction={laser beam recurs={30}{#1}},draw opacity=1,draw=#1},
}

\newcommand\Kfivedraw[5]{%
    \draw (#1) -- (#2) -- (#3) -- (#4) -- (#5) -- (#1);
    \draw (#1) -- (#3);
    \draw (#1) -- (#4);
    \draw (#2) -- (#4);
    \draw (#2) -- (#5);
    \draw (#3) -- (#5);
}

\tikzset{gradientpath/.style n args={3}{
    postaction={
    decorate,
    decoration={
    markings,
    mark=between positions 0 and \pgfdecoratedpathlength step 0.2pt with {
    \pgfmathsetmacro\myval{multiply(
        divide(
        \pgfkeysvalueof{/pgf/decoration/mark info/distance from start}, 
        \pgfdecoratedpathlength
        ),
        100
    )};
    \pgfsetfillcolor{#3!\myval!#2};
    \pgfpathcircle{\pgfpointorigin}{#1};
    \pgfusepath{fill};}
}}}}

% fpu reciprocal from https://tex.stackexchange.com/a/537016, seemingly helps avoid 'dimension too large' errors
\begin{tikzpicture}[use fpu reciprocal]

\begin{scope}[shift={(-4.0, 0.0)}, every node/.style={thick, circle, draw, minimum size=2.4mm, fill=white}]

\begin{scope}[scale=2.4]

\node[draw=none, label={[label distance=0.4cm]197.5:$v_i^j$}] (vij) at (0.0, 0.0) {};
\node[draw=none, label={[label distance=0.4cm]0:$c_i^j$}] (cij) at (1.0, 0.0) {};
\node[label={[label distance=0.4cm]180:$u_i^{j+1}$}] (uij1) at (0.0, -1.0) {};
\node[label={[label distance=0.4cm]0:$h_i^j$}] (hij) at (0.5, -0.5) {};
\node[draw=none, label={[label distance=0.4cm]0:$d_i^j$}] (dij) at (1.0, -1.0) {};

\begin{scope}[rotate=35]
\node[draw=none, label={[label distance=0.4cm]35:$b_i^j$}] (bij) at (1.0, 0.0) {};
\node[draw=none, label={[label distance=0.4cm]215:$u_i^j$}] (uij) at (0.0, 1.0) {};
\node[label={[label distance=0.4cm]35:$e_i^j$}] (eij) at (0.5, 0.5) {};
\node[draw=none, label={[label distance=0.4cm]35:$a_i^j$}] (aij) at (1.0, 1.0) {};
\begin{scope}[shift={(0.0, 1.0)}]
\begin{scope}[rotate=35]
\node[draw=none] (vijminus1) at (0.0, 1.0) {};
\node[draw=none] (hijminus1) at (0.5, 0.5) {};
\node[draw=none] (dijminus1) at (1.0, 0.0) {};
\node[draw=none] (cijminus1) at (1.0, 1.0) {};
\end{scope}
\end{scope}
\end{scope}

\begin{scope}[shift={(0.0, -1.0)}]
\begin{scope}[rotate=-35]
\node[draw=none, label={[label distance=0.4cm]-35:$a_i^{j+1}$}] (aij1) at (1.0, 0.0) {};
\node (vij1) at (0.0, -1.0) {};
\node[label={[label distance=0.4cm]-35:$e_i^{j+1}$}] (eij1) at (0.5, -0.5) {};
\node[label={[label distance=0.4cm]-35:$b_i^{j+1}$}] (bij1) at (1.0, -1.0) {};

\begin{scope}[shift={(0.0, -1.0)}]
\begin{scope}[rotate=-35]
\node[draw=none] (cij1) at (1.0, 0.0) {};
% \node (uij2) at (0.0, -1.0) {};
\node[draw=none] (hij1) at (0.5, -0.5) {};
% \node (bij12) at (1.0, -1.0) {};
\end{scope}
\end{scope}
\end{scope}
\end{scope}

\draw (aij) -- (bij) -- (vij) -- (uij) -- (aij);
\draw (vij) -- (cij) -- (dij) -- (hij);
\draw (aij) -- (eij) -- (bij);
\draw (uij) -- (eij) -- (vij);

\draw (vij) -- (hij) -- (cij);
\draw (eij) -- (hij);
\draw (bij) -- (cij);

\draw (bij) -- (hij);
\draw (eij) -- (cij);

\draw plot [smooth] coordinates {(aij) ($ (eij) !.4! (uij) $) (vij)};
\draw plot [smooth] coordinates {(bij) ($ (eij) !.4! (vij) $) (uij)};

\draw (aij) -- (hijminus1);

\draw (vij) -- (uij1);
\draw (dij) -- (aij1);
\draw (hij) -- (uij1);
\draw (dij) -- (uij1);
\draw (vij1) -- (uij1) -- (eij1) -- (vij1) -- (bij1) -- (eij1) -- (aij1) -- (bij1);

\draw plot [smooth] coordinates {(aij1) ($ (eij1) !.4! (uij1) $) (vij1)};
\draw plot [smooth] coordinates {(bij1) ($ (eij1) !.4! (vij1) $) (uij1)};

\draw (eij) -- (hijminus1);
\draw (eij) -- (dijminus1);

\draw (aij) -- (dijminus1);
\draw (uij) -- (vijminus1);
\draw (uij) -- (vijminus1);
\draw (uij) -- (dijminus1);
\draw (uij) -- (hijminus1);

\draw (hij) -- (eij1);
\draw (dij) -- (eij1);
\draw (hij) -- (aij1);
\draw (uij1) -- (aij1);

\draw (eij1) -- (hij1);
\draw (eij1) -- (cij1);

\draw (bij1) -- (hij1);
\draw (bij1) -- (cij1);

\draw [smooth] plot coordinates {(cij) ($ (hij) !.4! (vij) $) (uij1)};
\draw [smooth] plot coordinates {(dij) ($ (hij) !.4! (uij1) $) (vij)};
\draw [smooth] plot coordinates {(uij) ($ (hijminus1) !.4! (vijminus1) $) (cijminus1)};

\node (aijfake) at (aij) {};
\node (bijfake) at (bij) {};
\node (cijfake) at (cij) {};
\node (dijfake) at (dij) {};
\node (uijfake) at (uij) {};
\node (vijfake) at (vij) {};
\node (uij1fake) at (uij1) {};
\node (aij1fake) at (aij1) {};
\node (bij1fake) at (bij1) {};
\node (vij1fake) at (vij1) {};
\node (dijminus1fake) at (dijminus1) {};

\begin{scope}[rotate=17.5]
\begin{scope}[shift={(1.9, 0.0)}]
\node[label={[label distance=0.4cm]-72.5:$s_r^1$}] (sr1) at (0.0, -0.3) {};
\node[label={[label distance=0.4cm]107.5:$t_r^1$}] (tr1) at (0.0, 0.3) {};

\node[label={[label distance=0.4cm]107.5:$w_r^1$}] (wr1) at (0.4, 0.7) {};

\node[label={[label distance=0.4cm]-72.5:$w_r^2$}] (wr2) at (0.8, -0.3) {};
\node[label={[label distance=0.4cm]107.5:$w_r^3$}] (wr3) at (0.8, 0.3) {};

\node[label={[label distance=0.4cm]107.5:$w_r^4$}] (wr4) at (1.2, 0.7) {};

\node[label={[label distance=0.4cm]-72.5:$s_r^2$}] (sr2) at (1.6, -0.3) {};
\node[label={[label distance=0.4cm]107.5:$t_r^2$}] (tr2) at (1.6, 0.3) {};

\Kfivedraw{sr1}{tr1}{wr1}{wr2}{wr3}
\draw (wr3) -- (wr4) -- (sr2) -- (tr2) -- (wr2) -- (sr2) -- (wr3) -- (tr2) -- (wr4) -- (wr2);

% \shade[top color=red, path fading=we] (aij.center) -- ($(aij.center) + (-0.15, 0.7)$) -- ($(uij.center) + (-0.6, 0.3)$) -- (uij.center) -- cycle;

\end{scope}
\end{scope}
\end{scope}

\end{scope}

% do the shading nonsense
\begin{scope}
\path [laser beam=white] ($(aij.center) + (-0.2, 1.0)$) -- ($(uij.center) + (-2.0, -0.3)$);
\fill [white] ($(aij.center) + (-0.2, 1.0)$) -- ($(uij.center) + (-2.0, -0.3)$) -- ($(uij.center) + (-2.5, -0.0)$) -- ($(aij.center) + (-4.0, 2.0)$);

\path [laser beam=white] ($(uij1.center) + (-1.0, -0.4)$) -- ($(bij1.center) + (-0.9, -0.9)$);
\fill [white] ($(uij1.center) + (-1.0, -0.4)$) -- ($(uij1.center) + (-2.0, -0.5)$) -- ($(bij1.center) + (-3.0, -0.9)$) -- ($(bij1.center) + (-1.0, -1.0)$);
\end{scope}

\end{tikzpicture}

% \vspace*{-1.8cm}

%% file: sections/conclusion.tex
To recap, we considered two problems that involve finding a maximum-cardinality $K_r$-packing in an undirected graph of fixed maximum degree $\Delta$. In the first problem (\vdkr), the $K_r$-packing must be vertex disjoint. In the second problem (\edkr), it must be edge disjoint. It is known that \vdkthree is solvable in linear time if $\Delta=3$ but $\APX$-hard if $\Delta \geq 4$, and \edkthree is solvable in linear time if $\Delta=4$ but $\APX$-hard if $\Delta \geq 5$~\cite{caprara_packing_2002}. We generalised these results and presented a full complexity classification for both \vdkr and \edkr.

Specifically, we first showed that both \vdkr and \edkr are solvable in linear time if $\Delta < 3r/2 - 1$ (Theorem~\ref{thm:krpacking_vdkr_3r2minus1} and Corollary~\ref{cor:krpacking_edkr_3r2minus2}). We then showed that both \vdkr and \edkr are solvable in polynomial time if $\Delta < 5r/3 - 1$ (Theorems~\ref{thm:krpacking_vdkr_polytime_5r3minus1} and~\ref{thm:krpacking_edkrgeq6_polytime}). We then showed that if $r \leq 5$ then \edkr is actually solvable in polynomial time in the slightly more general case in which $\Delta \leq 2r - 2$ (Theorem~\ref{thm:krpacking_edkr345_polytime}). We then showed that \vdkr is $\APX$-hard if $r \geq 3$ and $\Delta \geq \lceil 5r/3 \rceil - 1$ (Theorem~\ref{thm:krpacking_vdkr_apxhard}) and \edkr is $\APX$-hard if either $r \geq 6$ and $\Delta \geq \lceil 5r/3 \rceil - 1$ (Theorem~\ref{thm:krpacking_edkr_rgeq6_apxhard}), or $3 \leq r \leq 5$ and $\Delta > 2r - 2$ (Theorem~\ref{thm:krpacking_edkr345apxhard}).

Some of our polynomial-time algorithms involved finding a maximum independent set in a corresponding intersection graph. In each case, we showed that this intersection graph was claw-free, from which it follows that a maximum independent set in the intersection graph can be found in polynomial time \cite{MINTY1980284,SBIHI198053}. As we noted in 
Section~\ref{sec:krpacking_ptime}, in a more general setting in which the vertices of a claw-free graph have real weights, it is possible to find an independent set of maximum weight in polynomial time~\cite{MINTY1980284,Nakamura01}. We remarked that our polynomial-time solvability results involving claw-free graphs can therefore be generalised to versions of \vdkr and \edkr in which vertices or edges have weights and the goal is to find a $K_r$-packing of maximum total weight. An interesting direction for future work could be to consider other weighted versions of \vdkr and \edkr. More generally, it might be interesting to explore whether there are other types of packing problem in which a ``natural'' condition implies that the intersection graph is claw-free.

Another direction for future work involves approximation algorithms. For example, Mani\'c and Wakabayashi~\cite{Eurocomb05} showed that the known approximation ratio of $(3/2 + \varepsilon)$ for \vdkthree and \edkthree with can be improved upon in the restricted settings where $\Delta = 4$ and $\Delta = 5$, respectively. It might be possible to show a similar improvement of the corresponding approximation ratio for \vdkr and \edkr in the setting of an arbitrary fixed maximum degree.

Another possible direction is parameterised complexity. If $r$ is not a fixed constant then our linear-time algorithms could be seen as $\FPT$ algorithms and our other polynomial-time algorithms could be seen as $\XP$ algorithms, in both cases relative to parameter $r$. It might be interesting to explore parameterised hardness with respect to parameter $r$, for values of $r$ where only $\XP$ algorithms are currently known.

% The second direction is to consider other problems involving clique packing. For example, Chataigner et al.~\cite{CHATAIGNER20091396} study the approximability of the $\mathcal{K}_r$-packing problem, in which the goal is to find a vertex-disjoint set of cliques, each with size \emph{at most} $r$, that covers the maximum number of edges in the input graph.

%% file: sections/funding.tex
This work was supported by the Engineering and Physical Sciences Research Council (grant numbers EP/R513222/1 and EP/X013618/1). The authors thank the anonymous reviewer for their detailed review and observation regarding a weighted variant of \vdkr.